\documentclass[aps,showpacs]{revtex4}

\usepackage{graphics,epsfig}
\usepackage{amssymb,revsymb}
\usepackage{bm}

\newcommand \A {A}
\newcommand \etaa {2\alpha_0}

\begin{document}
\title{Phase field crystal dynamics for binary systems: Derivation
  from dynamical density functional theory, amplitude equation
  formalism, and applications to alloy heterostructures}
\author{Zhi-Feng Huang}
\affiliation{Department of Physics and Astronomy, Wayne State
  University, Detroit, MI 48201}
\author{K. R. Elder}
\affiliation{Department of Physics, Oakland University, Rochester,
  MI 48309}
\author{Nikolas Provatas}
\affiliation{Department of Materials Science and Engineering
and Brockhouse Institute for Materials Research, McMaster
University, Hamilton, ON, Canada L8S-4L7}

\date{\today}

\begin{abstract}

The dynamics of phase field crystal (PFC) modeling is derived from
dynamical density functional theory (DDFT), for both single-component
and binary systems. The derivation is based on a truncation up to
the three-point direct correlation functions in DDFT, and the lowest order
approximation using scale analysis. The complete amplitude
equation formalism for binary PFC is developed to describe the
coupled dynamics of slowly varying complex amplitudes of structural
profile, zeroth-mode average atomic density, and system concentration
field. Effects of noise (corresponding to stochastic amplitude
equations) and species-dependent atomic mobilities are also
incorporated in this formalism. Results of a sample application to
the study of surface segregation and interface intermixing in alloy
heterostructures and strained layer growth are presented, showing the
effects of different atomic sizes and mobilities of alloy components.
A phenomenon of composition overshooting at the interface is found,
which can be connected to the surface segregation and enrichment of
one of the atomic components observed in recent experiments of alloying
heterostructures.

\end{abstract}

\pacs{
81.10.Aj, %Theory and models of crystal growth; physics and
          %chemistry of crystal growth, crystal morphology, and orientation
05.70.Ln, %Nonequilibrium and irreversible thermodynamics
81.15.Aa, %Theory and models of film growth
64.60.My  %Metastable phases
}

\maketitle

\section{Introduction}
\label{sec:intro}

Understanding the formation and dynamics of complex spatial structures
or patterns has been of continuing interest due to the 
fundamental importance of predicting and controlling system
properties and material functions. However, a comprehensive
understanding is hindered by the fact that the processes involved are
usually nonlinear, nonequilibrium, can span a variety of length and
time scales, and are highly influenced by the complex coupling with
materials growth and processing conditions. Typical examples include
the growth of strained solid films and the formation of
nanostructures such as quantum dots or nanowires, which involve the
interplay among microscopic crystalline structure, mesoscopic or
nanoscale surface pattern, topological defects (e.g., dislocations),
as well as various growth parameters such as temperature, misfit
strain, growth rate, and film thickness
\cite{re:politi00,re:stangl04,re:huang08}.
The system dynamics and evolution are further complicated in alloy
samples, due to the additional coupling to spatial/temporal variation
of alloy composition particularly in the case of phase separation
\cite{re:gunton83,re:guyer95}.

To address these complex phenomena a variety of theoretical modeling
and simulation methods have been developed, which can be roughly
characterized via the level of description that they focus on. At the
microscopic level capturing crystalline details, atomistic modeling
techniques such as Monte Carlo (MC) or molecular dynamics (MD) have
been widely adopted. For example, nanostructure (e.g., islands/pits)
formation during strained film epitaxy has been studied via the
kinetic MC method incorporating elastic interaction and strain energy
\cite{re:nandipati06,re:lung05,re:baskaran10}, while detailed structure and
dynamics of crystal defects like grain boundaries and dislocations
have been simulated by MD \cite{re:wolf05,re:derlet09}. However,
the limitation of small length and time scales addressed in these
atomistic methods leads to large computational demands and hence the
restriction of system size and evolution time range that can be accessed. 
Such limitation can be overcome via continuum modeling methods,
including continuum elasticity theory used in strained film growth
\cite{re:spencer91,re:guyer95,re:huang02a,re:huang02b,re:huang03a,%
re:huang03b,re:tu07,re:desai10}
and the well-known phase field models that have been applied to a wide
range of areas such as crystal growth, nucleation, phase separation,
solidification, defect dynamics, etc.
\cite{re:elder94,re:elder01,re:kassner01,re:echebarria04,re:granasy06}.
These continuum approaches feature coarse-grained, long-wavelength
scales and diffusive time dynamics, but are not formulated for the
short-wavelength scales associated with microscopic crystalline
details.

To incorporate the advantages of these approaches and hence be able to
simultaneously model crystalline details on length and time scales of
experimental relevance, the phase field crystal (PFC) 
\cite{re:elder02,re:elder04,re:elder07} model and the related
amplitude representation
\cite{re:goldenfeld05,re:athreya06,re:athreya07,re:chan10,re:yeon10} 
were developed recently.
The PFC model incorporates the small length scales of the 
crystalline structure with diffusive time scales by describing 
the dynamics of the atomic number density field
$\rho$, a continuum field variable that is spatially periodic at
atomic length scales in crystalline state \cite{re:elder02,re:elder04}.
%as well as alloy concentration field $\psi$ \cite{re:elder07}. 
To alleviate the limitation imposed by the necessity of describing small 
length scales, an amplitude representation was developed to describe
slowly varying envelope or amplitude functions while  
maintaining the basic features of crystalline states, particularly
elasticity, plasticity and multiple crystal orientations. 
Both the original PFC and corresponding amplitude representation have 
be extended to binary alloys 
\cite{re:elder04b,re:elder07,re:elder10}.
In the binary case the amplitude representation describes the 
amplitude and phase of the density
field \cite{re:goldenfeld05,re:athreya06,re:athreya07,re:yeon10} and
also the concentration profile \cite{re:elder10}, which is assumed to vary on
``slow'' scales compared to atomic lattice spacing.  A wide range of
phenomena has been studied via this PFC method for both pure and
binary material systems, including solidification
\cite{re:elder07,re:elder10,re:teeffelen09}, grain nucleation and
growth \cite{re:goldenfeld05,re:backofen07,re:yeon10,re:tegze09},
phase segregation \cite{re:elder07,re:elder10}, quantum dot growth
during epitaxy \cite{re:huang08,re:elder10,re:wu09,re:huang10},
surface energy anisotropy \cite{re:wu07,re:majaniemi09}, formation and
melting of dislocations and grain boundaries
\cite{re:berry06,re:berry08b,re:mellenthin08,re:spatschek10},  
commensurate/incommensurate transitions \cite{re:achim06,re:ramos08}, 
sliding friction \cite{re:achim09,re:ramos10},
and glass formation \cite{re:berry08}. 
In addition, recent work has been conducted
to extend the modeling to incorporate 
%the vacancy dynamics \cite{re:chan09}, 
faster time scales associated with mechanical relaxation  
\cite{re:stefanovic06,re:stefanovic09}, and to develop new efficient
computational methods \cite{re:athreya07,re:cheng08,re:hirouchi09,re:tegze09b}.

	The PFC model can be connected to 
microscopic theory via classical density functional theory (DFT) of
freezing \cite{re:elder07,re:majaniemi07,re:majaniemi08,re:jaatinen09,re:teeffelen09}. 
It has been found that the PFC free energy functional can be derived
from classical DFT for either pure materials or binary mixtures 
\cite{re:ramakrishnan79,re:haymet81,re:rick90,re:denton90,re:singh91,re:lowen94}, 
by approximating the two-point direct correlation function with a truncated 
Fourier series and expanding the ideal-gas part of the DFT free energy functional 
in a power series of $\rho$ and $\psi$ (up to 4th order)
\cite{re:elder07}.  While this connection provides insight into the 
parameters that enter PFC models,
the approximations used are quite drastic and the resulting model is 
a poor approximation of classical DFT \cite{re:jaatinen09}.   A similar 
connection could be made with the atomic density formulation of 
Jin and Khachaturyan \cite{re:jin06} which is similar in form to 
the classical DFT, although the parameters that enter are given a 
different physical interpretation.  

The main difficulty in directly simulating classical DFT is that the 
solutions for $\rho$ are very sharply peaked around the
lattice positions (at least in metallic crystals), while
simple PFC models predict very smooth sinusoidal profiles.  This difference 
makes numerical simulations of a simple PFC model much simpler than 
classical DFT as the former model's grid spacing can be a factor of ten larger than 
the latter's, so that in three dimensions a PFC model can simulate systems three orders of
magnitudes larger than classical DFT with the same memory requirements.   
In addition it has been shown that a simple PFC can be adjusted 
to match many material properties, such as surface energy and its
anisotropy, bulk moduli, and the miscibility gap 
in three-dimensional (3D) bcc iron \cite{re:jaatinen09} and the velocity of 
liquid/solid fronts in two-dimensional (2D) 
hexagonal crystal of colloids \cite{re:teeffelen09}.

Another benefit of PFC modeling is the ability to efficiently simulate microstructure dynamics. 
At present, PFC dynamics has been largely introduced phenomenologically using
time-dependent Ginzburg-Landau type dynamics \cite{re:elder02,re:elder07}.
Recent progress includes the derivation of hydrodynamic evolution
equations for crystalline solids based on the Poisson bracket
formalism and the simplification to PFC equations
\cite{re:majaniemi07,re:majaniemi08}. Very recently research has been
conducted to connect the PFC-type models with microscopic dynamics
(Smoluchowski equation) via dynamical density functional theory (DDFT)
\cite{re:teeffelen09}. These results were also based on the truncation
of DFT free energy up to two-point correlation function, and for
single-component systems.

In this paper we provide a systematic derivation of PFC dynamics from
DDFT, for both single-component and binary systems that involve the
evolution of atomic number density and alloy concentration fields 
(see Sec. \ref{sec:ddft}).
Our derivation includes contributions from three-point direct correlation functions, 
which have been shown important for the DFT calculations
\cite{re:haymet83,re:curtin88}. The original PFC models can be
recovered via the lowest order approximation of our DDFT results, with
the PFC parameters being connected to quantities of DFT correlation
functions. Our calculations can be directly extended to incorporate
fourth and higher order correlation functions in DFT.

To complete the PFC methodology for binary systems, the full
amplitude equation formalism is established for a 2D system with 
hexagonal/triangular crystalline symmetry. It incorporates the
effects of species-dependent atomic mobility and average (zeroth-mode)
atomic density that are usually coupled with the dynamics of structural
amplitudes and concentration field during system evolution but absent
in previous studies of binary PFC. As shown in Sec. \ref{sec:ampl},
the standard multiple-scale expansion is first applied to derive the
lowest order amplitude equations, followed by a hybrid approach that
we develop here to obtain the equations incorporating all orders of
expansion.
Furthermore, stochastic amplitude equations are derived for both
single-component and binary PFC models, showing the corresponding
noise dynamics as well as its coupling due to different atomic
mobilities of system components (see Sec. \ref{sec:noise}).

As has been discussed in previous research, the advantage of the
amplitude equation representation can be revealed via its large
increase of computational efficiency due to the large length and time
scales involved \cite{re:goldenfeld05,re:athreya06,re:yeon10} and also
its amenability to advanced numerical schemes such as adaptive mesh
refinement method \cite{re:athreya07}. Furthermore, these amplitude
equations are more amenable to analytic calculations as shown in recent
studies of surface nanostructure formation in strained epitaxial films
\cite{re:huang08,re:huang10} as well as in most recent results for
establishing the correspondence between PFC type models and
traditional phase field approaches \cite{re:elder10}. To further
illustrate these advantages, in Sec. \ref{sec:appl} we present a
sample application of the derived binary PFC amplitude equations to
the phenomenon of surface segregation and alloy intermixing. This is
of particular importance in material growth (e.g., group VI or III-V
semiconductor thin film epitaxy \cite{re:moison89,re:walther01,%
re:cederberg07,re:walther97,re:denker05,re:gerling01,re:dorin03,re:pearson04}),
but rather limited information and understanding is available to date.
We focus on both liquid-solid(crystal) coexistence profile and the
coherent growth of strained solid layers, and show the control of
intra- and inter-layer diffusion by varying material parameters
including solute expansion coefficient (due to different atomic
sizes), misfit strain in alloy layers, and the mobility
difference between alloy components.  This study provides an
understanding of mass transport mechanisms during material growth and
evolution. The dynamic processes of strained layer growth as well as
the associated composition overshooting phenomenon are obtained in our
calculations in Sec. \ref{sec:appl}.
The results are compared to experimental findings of
vertical composition segregation or surface enrichment as widely
encountered during the growth of various alloy heterostructure systems
such as InAs/GaAs, Ge/Si, GaAs/GaSb, InP/InGaAs,
etc. \cite{re:moison89,re:walther01,re:cederberg07,re:walther97,%
re:denker05,re:gerling01,re:dorin03,re:pearson04}.

\section{Derivation of PFC dynamics via dynamical density functional
  theory}
\label{sec:ddft}

\subsection{Single-component systems}
\label{sec:ddft_pure}

We start from the DDFT equation governing the evolution of a
time-dependent local atomic number density field $\rho({\bm r},t)$,
\begin{equation}
\frac{\partial \rho({\bm r},t)}{\partial t} = {\bm \nabla} \cdot \left [ 
M \rho({\bm r},t) {\bm \nabla} \frac{\delta {\cal F}}{\delta \rho} \right ],
\label{eq:ddft}
\end{equation}
which was first proposed phenomenologically
\cite{re:evans79,re:dieterich90} and was later derived  
by various groups via microscopic Brownian dynamics
\cite{re:marconi99,re:archer04a,re:marconi08} and Hamiltonian dynamics 
and hydrodynamics \cite{re:chan05}
(see Ref. \cite{re:teeffelen09} for a brief review).
The DDFT equations for binary A/B systems are similar to
Eq. (\ref{eq:ddft}), with $\rho({\bm r},t)$ replaced by
$\rho_i({\bm r},t)$ ($i=A,B$; see Sec. \ref{sec:ddft_binary}
below), which has also been derived recently from Brownian dynamics
(the Smoluchowski equation) \cite{re:archer05a}.
In Eq. (\ref{eq:ddft}) the mobility is $M=D/k_BT$, where $D$ is the diffusion
coefficient and $T$ is temperature.  

In classical DFT the free energy functional ${\cal F}$ can be expanded as
\cite{re:ramakrishnan79,re:haymet81}
\begin{eqnarray}
\frac{{\cal F}[\rho]}{k_BT} &=& \int d{\bm r} \left[ \rho \ln (\rho / \rho_l) -
  \delta \rho \right] - \frac{1}{2!} \int d{\bm r_1} d{\bm r_2} \delta
\rho({\bm r_1}) C^{(2)}({\bm r_1}, {\bm r_2}) \delta \rho({\bm r_2}) \nonumber\\
&-&\frac{1}{3!} \int d{\bm r_1} d{\bm r_2} d{\bm r_3} C^{(3)}({\bm r_1},
{\bm r_2}, {\bm r_3}) \delta \rho({\bm r_1}) \delta \rho({\bm r_2})
\delta \rho({\bm r_3}) + \cdots,
\label{eq:F_rho_C3}
\end{eqnarray}
where $\rho_l$ is the reference liquid state density taken at 
liquid/solid coexistence, $\delta \rho =
\rho - \rho_l$ and $C^{(n)}$ is the $n$-point direct correlation function of
the liquid phase at $\rho_l$. It is important that the correlation functions are taken 
from the liquid state to maintain rotational invariance.
Details of the correlation functions depend on the
specific material systems studied and are usually calculated via
various approximations \cite{re:rick90,re:singh91,re:lowen94}.
Following the original PFC approach \cite{re:elder07}, the Fourier
component of the two-point correlation function $\hat{C}^{(2)}$ is
expanded as a power series of wavenumber $q$ to fit up to its first
peak, i.e.,
\begin{equation}
\hat{C}^{(2)}(q)=-\hat{C}_0+\hat{C}_2 q^2-\hat{C}_4 q^4+\cdots,
\label{eq:C2}
\end{equation}
where $\hat{C}_0$, $\hat{C}_2$, and $\hat{C}_4$ are fitting parameters
that can be connected to material properties such as isothermal
compressibility of liquid phase, bulk modulus and lattice constant of
crystal state \cite{re:elder07,re:jaatinen09}. For the three-point
correlation function $C_3$, its Fourier transform yields
$$
C^{(3)}({\bm r_1}, {\bm r_2}, {\bm r_3}) = \frac{1}{(2\pi)^6}
\int d{\bm q} d{\bm q'} e^{i{\bm q} \cdot ({\bm r_1}-{\bm r_2})}
 e^{i{\bm q} \cdot ({\bm r_2}-{\bm r_3})} \hat{C}^{(3)}({\bm q}, 
{\bm q'}).
$$
The simplest approximation is to keep only the zero wavenumber mode,
i.e.,
\begin{equation}
\hat{C}^{(3)}({\bm q},{\bm q'}) \simeq \hat{C}^{(3)}({\bm q}={\bm
  q'}=0)=-\hat{C}_0^{(3)},
\label{eq:C3}
\end{equation}
as adopted in the DFT studies of hard spheres
\cite{re:haymet83,re:smithline87} and Lennard-Jones mixtures
\cite{re:rick89}. This can be justified from the previous results of
hard-spheres DFT calculations that nonzero wavenumber components of
$\hat{C}^{(3)}$ have been shown to yield minor contributions
\cite{re:haymet83,re:smithline87} and that as order $n$ increases, the
oscillation details of $\hat{C}^{(n)}$ become less and less relevant
compared to the zero wavenumber mode \cite{re:zhou00}.

Defining the rescaled atomic density field $n=(\rho - \rho_l)/\rho_l$
and using the approximations (\ref{eq:C2}) and (\ref{eq:C3}), the free
energy functional (\ref{eq:F_rho_C3}) becomes
\begin{equation}
\Delta {\cal F}/\rho_l k_B T = \int d {\bm r} \left [ (1+n)\ln(1+n)
+ \frac{1}{2} B^x n \left ( 2R^2 \nabla^2 + R^4 \nabla^4 \right ) n
+ \frac{1}{2} B_l' n^2 + \frac{1}{3} \tilde{B} n^3 \right ],
\label{eq:F_C3}
\end{equation}
where $\Delta {\cal F} = {\cal F} [\rho] - {\cal F}[\rho_l]$, and
\begin{equation}
B_l'=\rho_l \hat{C}_0=B^{\ell}-1, \qquad B^x = \rho_l \hat{C}_2^2 / 4\hat{C}_4,
\qquad R=\sqrt{2\hat{C}_4/\hat{C}_2}, \qquad 
\tilde{B} = \rho_l^2 \hat{C}_0^{(3)}/2.
\end{equation}
Substituting Eq. (\ref{eq:F_C3}) into the DDFT equation
(\ref{eq:ddft}), which can be rewritten as
\begin{equation}
\frac{\partial n}{\partial t} = M' {\bm \nabla} \cdot \left [ (1+n) {\bm
    \nabla} \frac{\delta {\cal F}}{\delta n} \right ]
\label{eq:ddft_n}
\end{equation}
(with $M'=M/\rho_l$), we find \cite{re:notes_pfc}
\begin{equation}
\frac{\partial n}{\partial t} = D \left \{ \nabla^2  \left [ 
-( B^x - B^{\ell} ) n + B^x \left ( R^2 \nabla^2 + 1 \right )^2
n + \tau n^2 + v n^3 \right ]
+ B^x {\bm \nabla} \cdot \left [ n {\bm \nabla}
  \left ( R^2 \nabla^2 + 1 \right )^2 n \right ] \right \},
\label{eq:pfc}
\end{equation}
where $\tau = -(B^x-B^{\ell}+1)/2+\tilde{B}$, $v=2\tilde{B}/3$, and we have
used the relation $M =D/k_B T$. Note that if only the 
two-point correlation function in the DFT free energy
(\ref{eq:F_rho_C3}) was used it would yield $\tilde{B}=v=0$, and Eq. (\ref{eq:pfc}) 
reduces to a form equivalent to the PFC1 model given in
Ref. \cite{re:teeffelen09}. However, this would then be a 2nd-order
dynamic equation due to the absence of $n^3$ term, and as found in our
numerical tests, is more difficult to converge at long enough time
compared to the full 3rd-order Eq. (\ref{eq:pfc}). 

It is convenient to rescale Eq. (\ref{eq:pfc}) via a length scale
$R$, a time scale $R^2/DB^x$, and $n \rightarrow \sqrt{v/B^x}~ n$,
leading to
\begin{equation}
\frac{\partial n}{\partial t} = \nabla^2  \left [ -\epsilon n +
 \left ( \nabla^2 + q_0^2 \right )^2 n + g_2 n^2 + n^3 \right ]
+ g_0 {\bm \nabla} \cdot \left [ n {\bm \nabla} \left ( \nabla^2 +
    q_0^2 \right )^2 n \right ], \label{eq:pfc_re}
\end{equation}
where
\begin{equation}
\epsilon = (B^x - B^{\ell})/B^x, \quad q_0=1, \quad 
g_2 = \tau / \sqrt{v B^x}, \quad g_0 = \sqrt{B^x / v}.
\label{eq:para_pfc}
\end{equation}
The original PFC equation is recovered by considering that 
(${\bm \nabla} \cdot [ n {\bm \nabla} ( \nabla^2 + q_0^2 )^2 n]$) is
of higher order compared to term $\nabla^2 ( \nabla^2 + q_0^2 )^2
n$. This can be obtained via a simple scale analysis: 
$n \sim {\cal O}(\epsilon^{1/2})$ and $( \nabla^2 + q_0^2 )^2 n \sim
{\cal O}(\epsilon^{3/2})$ (see also Sec. \ref{sec:multiscale} for more
detail of scale expansion). Thus to lowest order approximation,
Eq. (\ref{eq:pfc_re}) can be reduced to the original PFC model
equation that has been widely used:
\begin{equation}
\frac{\partial n}{\partial t} = \nabla^2  \left [ -\epsilon n +
 \left ( \nabla^2 + q_0^2 \right )^2 n + g_2 n^2 + n^3 \right ].
\end{equation}

This derivation procedure can be readily extended to incorporate
higher order direct correlation functions of DFT (e.g., four-point,
five-point, etc.) and thus to include higher order terms such as
$n^4$, $n^5$, ..., in the PFC model. Similarly, these high-order
correlation functions can be effectively approximated to lowest order 
via the zero wavenumber modes, based on the recent DFT calculations
\cite{re:zhou00}.  For example, the contribution ${\cal F}^{(4)}$ of
free energy functional from the four-point correlation function is
given by
\begin{equation}
{\cal F}^{(4)}/k_BT = - \frac{1}{24} \int d{\bm r_1} d{\bm r_2} d{\bm r_3}
d{\bm r_4} C^{(4)}({\bm r_1}, {\bm r_2}, {\bm r_3}, {\bm r_4}) \delta
\rho({\bm r_1}) \delta \rho({\bm r_2}) \delta \rho({\bm r_3}) \delta
\rho({\bm r_4}).
\end{equation}
Assuming $\hat{C}^{(4)}({\bm q},{\bm q'},{\bm q''}) \simeq
\hat{C}^{(4)}({\bm q}={\bm q'}={\bm q''}=0)=-\hat{C}_0^{(4)}$, the
free energy functional (\ref{eq:F_C3}) is now
\begin{equation}
\Delta {\cal F}/\rho_l k_B T = \int d {\bm r} \left [ (1+n)\ln(1+n)
+ \frac{1}{2} B^x n \left ( 2R^2 \nabla^2 + R^4 \nabla^4 \right ) n
+ \frac{1}{2} B_l' n^2 + \frac{1}{3} \tilde{B} n^3
+ \frac{1}{4} \tilde{B}_4 n^4 \right ],
\label{eq:F_C4}
\end{equation}
where $\tilde{B}_4 = \rho_l^3 \hat{C}_0^{(4)}/6$. The dynamic equation
for $n$ would then be
\begin{equation}
\frac{\partial n}{\partial t} = D \left \{ \nabla^2  \left [ 
-( B^x - B^{\ell} ) n + B^x \left ( R^2 \nabla^2 + 1 \right )^2
n + \tau n^2 + v n^3 + u n^4 \right ]
+ B^x {\bm \nabla} \cdot \left [ n \left ( R^2 \nabla^2 + 1
    \right )^2 {\bm \nabla} n \right ] \right \},
\label{eq:pfc_C4}
\end{equation}
where $v=2\tilde{B}/3 + \tilde{B}_4$ and $u=3\tilde{B}_4/4$. Again the
last term of Eq. (\ref{eq:pfc_C4}) is of higher order and can be
neglected in the lowest order approximation.

\subsection{Binary systems}
\label{sec:ddft_binary}

For a binary system with components A and B, the DDFT equations
describing the dynamics of the respective atomic density fields
$\rho_A$ and $\rho_B$ are given by \cite{re:archer05a}
\begin{equation}
\frac{\partial \rho_A}{\partial t} = {\bm \nabla} \cdot \left [ 
M_A \rho_A {\bm \nabla} \frac{\delta {\cal F}}{\delta \rho_A}
\right ], \qquad
\frac{\partial \rho_B}{\partial t} = {\bm \nabla} \cdot \left [ 
M_B \rho_B {\bm \nabla} \frac{\delta {\cal F}}{\delta \rho_B}
\right ],
\label{eq:ddft_AB}
\end{equation}
where $M_{A(B)}$ is the atomic mobility for specie $A$ ($B$). The
corresponding classical density functional free energy (hereafter referred to as 
"DFT" for short) is of the form \cite{re:rick90,re:denton90,re:singh91,re:lowen94}
\begin{eqnarray}
&{\cal F}/k_BT =& \int d{\bm r} \sum_{i=A,B} \left [ \rho_i \ln
  \frac{\rho_i }{\rho_l^i} - \delta \rho_i \right ] \nonumber\\
&& - \sum_{n=2}^{\infty}\frac{1}{n!} \int d{\bm r_1} \cdots d{\bm r_n}
\sum_{i,...,j=A, B} C_{i...j}^{(n)}({\bm r_1}, \cdots, {\bm r_n}) \delta
\rho_i({\bm r_1}) \cdots \delta \rho_j ({\bm r_n}),
\end{eqnarray}
where $\rho_l^i$ is the reference liquid state density of component
$i$, $\delta \rho_i = \rho_i - \rho_l^i$, and $C_{i...j}^{(n)}$ refers
to the $n$-point direct correlation function between components
$i,...,j=A, B$. Up to three-point correlation functions, we have
\begin{eqnarray}
{\cal F}/k_BT &=& \int d{\bm r} \left [ 
\rho_A \ln(\rho_A/\rho_l^A)-\delta \rho_A
+\rho_B \ln(\rho_B/\rho_l^B)-\delta \rho_B \right ] \nonumber\\
&-& \frac{1}{2} \int d{\bm r_1} d{\bm r_2} \left [ \delta 
\rho_A({\bm r_1}) C_{AA}^{(2)}({\bm r_1}, {\bm r_2}) \delta \rho_A({\bm r_2}) 
+ \delta \rho_B({\bm r_1}) C_{BB}^{(2)}({\bm r_1}, {\bm r_2}) 
\delta \rho_B({\bm r_2}) 
+ 2 \delta \rho_A({\bm r_1}) C_{AB}^{(2)}({\bm r_1}, {\bm r_2}) \delta
\rho_B({\bm r_2}) \right ] \nonumber\\
&-&\frac{1}{6} \int d{\bm r_1} d{\bm r_2} d{\bm r_3} \left [
C_{AAA}^{(3)}({\bm r_1},{\bm r_2}, {\bm r_3}) \delta \rho_A({\bm r_1})
\delta \rho_A({\bm r_2}) \delta \rho_A({\bm r_3}) 
+ C_{BBB}^{(3)}({\bm r_1},{\bm r_2}, {\bm r_3}) \delta \rho_B({\bm r_1})
\delta \rho_B({\bm r_2}) \delta \rho_B({\bm r_3}) \right. \nonumber\\
&& \left. + 3C_{AAB}^{(3)}({\bm r_1},{\bm r_2}, {\bm r_3}) 
\delta \rho_A({\bm r_1}) \delta \rho_A({\bm r_2}) \delta \rho_B({\bm r_3})
+ 3C_{ABB}^{(3)}({\bm r_1},{\bm r_2}, {\bm r_3}) \delta \rho_A({\bm r_1})
\delta \rho_B({\bm r_2}) \delta \rho_B({\bm r_3}) \right ].
\label{eq:F_AB}
\end{eqnarray}
Similar to the single-component case discussed in
Sec. \ref{sec:ddft_pure}, the correlation functions $C_{ij}^{(2)}
({\bm r_1}, {\bm r_2})$ and $C_{ijk}^{(3)}({\bm r_1},{\bm r_2},
{\bm r_3})$ ($i,j,k=A,B$) are expanded in Fourier space as
\begin{eqnarray}
&\hat{C}_{ij}^{(2)}(q)=-\hat{C}_0^{ij}+\hat{C}_2^{ij} q^2-\hat{C}_4^{ij} q^4
+\cdots,& \nonumber\\
&\hat{C}_{ijk}^{(3)}({\bm q},{\bm q'}) \simeq \hat{C}_{ijk}^{(3)}
({\bm q}={\bm q'}=0)=-\hat{C}_0^{ijk}.&
\label{eq:C_expan}
\end{eqnarray}

As in the original binary PFC model, we introduce an atomic density
field $n$ and a concentration field $\psi$ via
\begin{equation}
n = \frac{\rho - \rho_l}{\rho_l} = \frac{\rho_A + \rho_B -
    \rho_l}{\rho_l}, \qquad
\psi = \frac{\rho_A - \rho_B}{\rho} = \frac{\rho_A -
  \rho_B}{\rho_A + \rho_B},
\label{eq:ndN}
\end{equation}
where $\rho_l = \rho_l^A + \rho_l^B$, and hence
\begin{equation}
\rho_A = \frac{\rho_l}{2} (1+n)(1+\psi), \qquad
\rho_B = \frac{\rho_l}{2} (1+n)(1-\psi).
\label{eq:rho_AB}
\end{equation}
Substituting Eqs. (\ref{eq:C_expan})--(\ref{eq:rho_AB}) into
(\ref{eq:F_AB}), we can express the free energy functional in terms of
$n$ and $\psi$:
\begin{eqnarray}
&\Delta {\cal F}/\rho_l k_BT =& \int d{\bm r} \left \{
(1+n)\ln(1+n) + \frac{1}{2} (1+n) \left [ (1+\psi) \ln(1+\psi) 
+(1-\psi) \ln(1-\psi) \right ] \right. \nonumber\\
&& + \beta(\psi) n + \frac{1}{2} B_l'(\psi) n^2
+ \frac{1}{3} \tilde{B}(\psi) n^3 + \frac{1}{2} \beta_2 \psi^2 
+ \frac{1}{3} \beta_3 \psi^3 \label{eq:F_ndN} \\
&& \left. + \frac{(1+n)}{2} \left ( 2B^x(\psi) R^2 \nabla^2 + B^x(\psi) R^4
\nabla^4 \right ) n + \frac{K}{2} \left | {\bm \nabla} [(1+n)\psi]
\right |^2 + \frac{\kappa}{2} \left ( \nabla^2 [(1+n)\psi]
\right )^2 \right \}, \nonumber
\label{eq:freeC3}
\end{eqnarray}
where 
\begin{eqnarray}
\beta (\psi) &=& \beta_0 \psi + \beta_1 \psi^2 
+ \beta_3 \psi^3 \nonumber\\
&=& \frac{\rho_l}{4} \left [ \delta \hat{C}_0 + \frac{\rho_l^B -
    \rho_l^A}{2} \delta \hat{C}_0^{(3)} \right ] \psi 
+ \left [ \beta_2 + \frac{\rho_l^2}{16} \delta \hat{C}_0^{(3)} \right
] \psi^2 + \frac{\rho_l^2}{16} \Delta \hat{C}_0^{(3)}
\psi^3, \nonumber\\
B_l'(\psi) &=& B^{\ell}(\psi) -1 
= B_0^{\ell}-1 + B_1^{\ell} \psi + B_2^{\ell} \psi^2 + B_3^{\ell} \psi^3
\nonumber\\
&=& \rho_l \left [ \hat{\bar{C}}_0 + \frac{\rho_l^B-\rho_l^A}{8}
  \hat{\tilde{C}}_0^{(3)} \right ] + \frac{\rho_l}{2} \left [
\delta \hat{C}_0 + \frac{\rho_l^B-\rho_l^A}{2} \delta \hat{C}_0^{(3)}
+ \frac{\rho_l}{4} \hat{\tilde{C}}_0^{(3)} \right ] \psi
\nonumber\\
&+& \frac{\rho_l}{4} \left [ \Delta \hat{C}_0 +
  \frac{\rho_l^B-\rho_l^A}{2} \Delta \hat{C}_0^{(3)} + \rho_l
  \delta \hat{C}_0^{(3)} \right ] \psi^2
+ \frac{\rho_l^2}{8} \Delta \hat{C}_0^{(3)} \psi^3,
\nonumber\\
\tilde{B}(\psi) &=& \frac{\rho_l^2}{16} \left [
  8\hat{\bar{C}}_0^{(3)} + 3\hat{\tilde{C}}_0^{(3)} \psi + 3\delta
  \hat{C}_0^{(3)} \psi^2 + \Delta \hat{C}_0^{(3)} \psi^3
\right ] = \tilde{B}_0 + \tilde{B}_1 \psi
+ \tilde{B}_2 \psi^2 + \beta_3 \psi^3, \nonumber\\
\beta_2 &=& \frac{\rho_l}{4} \left [ \Delta \hat{C}_0 + \frac{\rho_l^B -
    \rho_l^A}{2} \Delta \hat{C}_0^{(3)} \right ], \qquad
\Delta \beta = \beta_1 - \beta_2 = \frac{\rho_l^2}{16} \delta
\hat{C}_0^{(3)}, \qquad \tilde{B}_2 = 3\Delta \beta, \nonumber\\
B_1^{\ell} &=& 2 \beta_0 + \frac{2}{3} \tilde{B}_1, \qquad
B_2^{\ell} = 4\beta_1 - 3\beta_2, \qquad B_3^{\ell} = 2\beta_3, \nonumber\\
B^x(\psi) &=& \frac{\rho_l \left ( \hat{\bar{C}}_2 + \delta \hat{C}_2
\psi/2 \right )^2}{4\left ( \hat{\bar{C}}_4 + \delta \hat{C}_4
\psi/2 \right )} = \frac{\rho_l \hat{\bar{C}}_2^2}{4\hat{\bar{C}}_4}
\left [ 1 + \left ( \frac{\delta \hat{C}_2}{\hat{\bar{C}}_2} - 
\frac{\delta \hat{C}_4}{2\hat{\bar{C}}_4} \right ) \psi + \cdots
\right ] = B_0^x + B_1^x \psi + \cdots, \nonumber\\
R &=& \sqrt{\frac{2 \left ( \hat{\bar{C}}_4 + \delta \hat{C}_4 \psi/2
    \right )}{\hat{\bar{C}}_2 + \delta \hat{C}_2 \psi/2}}
= \sqrt{\frac{2\hat{\bar{C}}_4}{\hat{\bar{C}}_2}} \left [ 1 + 
\frac{1}{4} \left ( \frac{\delta \hat{C}_4}{\hat{\bar{C}}_4}
- \frac{\delta \hat{C}_2}{\hat{\bar{C}}_2} \right ) \psi + \cdots
\right ] = R_0 + R_1 \psi + \cdots, \nonumber\\
B^xR^2 &=& \frac{\rho_l}{2} \left ( \hat{\bar{C}}_2 + \frac{1}{2}
  \delta \hat{C}_2 \psi \right ) = B_0^xR_0^2 (1+\alpha_2 \psi),
\qquad \alpha_2 = \delta \hat{C}_2 / 2\hat{\bar{C}}_2,
\nonumber\\
B^xR^4 &=& \rho_l \left ( \hat{\bar{C}}_4 + \frac{1}{2}
  \delta \hat{C}_4 \psi \right ) = B_0^xR_0^4 (1+\alpha_4 \psi),
\qquad \alpha_4 = \delta \hat{C}_4 / 2\hat{\bar{C}}_4,
\nonumber\\
K &=& -\frac{\rho_l}{4} \Delta \hat{C}_2, \qquad
\kappa = \frac{\rho_l}{4} \Delta \hat{C}_4.
\label{eq:parameters}
\end{eqnarray}
In the above formulae, the following has been defined from the correlation functions:
\begin{eqnarray}
&\bar{C} = \frac{1}{4} \left ( C_{AA}^{(2)} + C_{BB}^{(2)} +
  2C_{AB}^{(2)} \right ),
\qquad \delta C = C_{AA}^{(2)} - C_{BB}^{(2)}, \qquad
\Delta C =  C_{AA}^{(2)} + C_{BB}^{(2)} - 2C_{AB}^{(2)}, & \nonumber\\
&\bar{C}^{(3)} = \frac{1}{8} \left ( C_{AAA}^{(3)} + C_{BBB}^{(3)}
+ 3C_{AAB}^{(3)} + 3C_{ABB}^{(3)} \right ), \qquad 
\tilde{C}^{(3)} = C_{AAA}^{(3)}
- C_{BBB}^{(3)} + C_{AAB}^{(3)} - C_{ABB}^{(3)}, \label{eq:CAB}\\
&\delta C^{(3)} = C_{AAA}^{(3)} + C_{BBB}^{(3)} - C_{AAB}^{(3)} 
- C_{ABB}^{(3)}, \qquad \Delta C^{(3)} = C_{AAA}^{(3)} - C_{BBB}^{(3)}
-3 C_{AAB}^{(3)} + 3 C_{ABB}^{(3)}, \nonumber
\end{eqnarray}
and the ``$\,\,\hat{}$\,\," in Eq.~(\ref{eq:parameters}) refer to the Fourier 
coefficients in the expansions of Eq. (\ref{eq:C_expan}) where the numerical
subscripts on the coefficients refer to the appropriate power
of the expansion.  For binary alloys the lattice
constant is often approximated by Vegard's law, i.e., 
$R \simeq R_0 + R_1 \psi = R_0 (1 + \alpha \psi)$.  In this expansion, near 
$\psi=0$ the solute expansion coefficient
$\alpha$ is expressed as
\begin{equation}
\alpha = R_1/R_0 = \frac{1}{2} (\alpha_4 - \alpha_2).
\label{eq:alpha}
\end{equation}
(In the dilute limit (i.e., $\psi \sim \pm 1$) it would be
simple to expand $R$ around $\psi \sim \pm 1$ to obtain the 
solute expansion coefficient as well.)
Using the simplification adopted in the original binary PFC
\cite{re:elder10}, it is assumed that $B^x \simeq B_0^x$ and
$R^2 \simeq R_0^2 (1+2\alpha \psi)$, $R^4 \simeq R_0^4(1+4\alpha
\psi)$ via expansion, which corresponds to the assumption of
$\alpha_2 \simeq 2\alpha$ and $\alpha_4 \simeq 4\alpha$ as obtained
from Eqs. (\ref{eq:parameters}) and (\ref{eq:alpha}).

In terms of the above definitions, the time derivatives of the
variables $n$ and $\psi$ defined in Eq. (\ref{eq:ndN}) are given by 
\begin{equation}
\frac{\partial n}{\partial t} = \frac{1}{\rho_l} \left (
\frac{\partial \rho_A}{\partial t}
+ \frac{\partial \rho_B}{\partial t}  \right ), \qquad
\frac{\partial \psi}{\partial t} = \frac{1}{\rho_l (1+n)} 
\left [ (1-\psi) \frac{\partial \rho_A}{\partial t}
- (1+\psi) \frac{\partial \rho_B}{\partial t} \right ].
\label{eq:ndN_rho}
\end{equation}
From the DDFT equations (\ref{eq:ddft_AB}), the dynamics 
for $n$ and $\psi$ thus become 
\begin{eqnarray}
&\partial n / \partial t =& M_1 {\cal D}_1 + M_2 {\cal D}_2,
\nonumber\\
&\partial \psi / \partial t =& \frac{1}{1+n} \left [
(M_2-M_1 \psi) {\cal D}_1 + (M_1-M_2 \psi) {\cal D}_2 
\right ], \label{eq:pfc_ndN}
\end{eqnarray}
where
\begin{equation}
M_1 = \frac{1}{2} k_BT \left ( M_A + M_B \right ), \qquad
M_2 = \frac{1}{2} k_BT \left ( M_A - M_B \right ),
\end{equation}
and
\begin{eqnarray}
&{\cal D}_1 =& \frac{1}{\rho_l k_BT} \left \{
 {\bm \nabla} \cdot \left [ (1+n) {\bm
    \nabla} \frac{\delta {\cal F}}{\delta n} \right ]
- {\bm \nabla} \cdot \left [ ({\bm \nabla} \psi) 
\frac{\delta {\cal F}}{\delta \psi} \right ] \right \},
\nonumber\\
&{\cal D}_2 =& \frac{1}{\rho_l k_BT} \left \{
{\bm \nabla} \cdot \left [ (1+n) \psi 
{\bm \nabla} \frac{\delta {\cal F}}{\delta n} \right ]
+ {\bm \nabla} \cdot \left [ (1+n)(1-\psi^2) {\bm \nabla}
\left ( \frac{1}{1+n} \frac{\delta {\cal F}}{\delta \psi} 
\right ) - (\psi {\bm \nabla} \psi) 
\frac{\delta {\cal F}}{\delta \psi} \right ] \right \}.
\label{eq:D_12}
\end{eqnarray}

Using the free energy functional (\ref{eq:F_ndN}) as well as
Eq. (\ref{eq:D_12}), the results of ${\cal D}_1$ and ${\cal D}_2$
(keeping all the terms) are
\begin{eqnarray}
&{\cal D}_1 =& 
\nabla^2 \left \{ - \left ( B_0^x-B_0^{\ell} \right ) n + 
\left ( B_1^{\ell} \psi + B_2^{\ell} \psi^2 \right ) n \right. \nonumber\\
&& + \left [ -(B_0^x-B_0^{\ell}+1)/2+\tilde{B}_0 
  + \left ( B_1^{\ell}/2+\tilde{B}_1
  \right ) \psi + \left ( B_2^{\ell}/2+\tilde{B}_2 \right ) \psi^2
\right ] n^2 \nonumber\\
&& + \frac{2}{3} \left ( \tilde{B}_0 + \tilde{B}_1 \psi
+ \tilde{B}_2 \psi^2 \right ) n^3 + \frac{1}{3} B_3^{\ell} (1+n)^3
\psi^3 + \beta_0 \psi + \frac{1}{2} (\beta_1+\Delta \beta)
\psi^2 \nonumber\\
&& \left. + B_0^x \left ( R_0^2\nabla^2 + 1 \right )^2 n + B_0^x \left ( 
\alpha_2 R_0^2\nabla^2 + \frac{\alpha_4}{2} R_0^4\nabla^4 \right )
[(1+n)\psi] \right \} \nonumber\\
&& + {\bm \nabla} \cdot \left \{ n {\bm \nabla} \left [
B_0^x \left ( R_0^2\nabla^2 + 1 \right )^2 n + B_0^x \left ( 
\alpha_2 R_0^2\nabla^2 + \frac{\alpha_4}{2} R_0^4\nabla^4 \right )
[(1+n)\psi] \right ] \right \} \nonumber\\
&& + {\bm \nabla} \cdot \left \{ (1+n)\psi {\bm \nabla} \left [ 
B_0^x \left ( \alpha_2 R_0^2\nabla^2 + \frac{\alpha_4}{2}
  R_0^4\nabla^4 \right ) n + \left ( -K\nabla^2 + \kappa \nabla^4
\right ) [(1+n)\psi] \right ] \right \}, \label{eq:D1}
\end{eqnarray}
\begin{eqnarray}
&{\cal D}_2 =& 
\nabla^2 \left \{ \beta_0 n + \left ( \beta_0/2 +
\tilde{B}_1/3 \right ) n^2 + \frac{2}{9} \tilde{B}_1 n^3 
+ (1+\beta_2+2\Delta \beta n) (1+n) \psi  \right. \nonumber\\
&& + \frac{1}{2}(\beta_0+B_3^{\ell}) (1+n)^2 \psi^2
 + \frac{1}{3}\tilde{B}_1 n^2 \psi^2 + \frac{2}{3} \Delta \beta
 (1+n)^3 \psi^3 \nonumber\\
&& \left. + B_0^x \left ( \alpha_2 R_0^2\nabla^2 + \frac{\alpha_4}{2}
  R_0^4\nabla^4 \right ) n + \left ( -K\nabla^2 + \kappa \nabla^4
\right ) [(1+n)\psi] \right \} \nonumber\\
&& + {\bm \nabla} \cdot \left \{ n {\bm \nabla} \left [
(1+n)(\beta_2+2\Delta \beta n) \psi + \beta_3 (1+n)^2 \psi^2
+ B_0^x \left ( \alpha_2 R_0^2\nabla^2 + \frac{\alpha_4}{2}
  R_0^4\nabla^4 \right ) n \right. \right. \nonumber\\
&& \left. \left. + \left ( -K\nabla^2 + \kappa \nabla^4
\right ) [(1+n)\psi] \right ] \right \}
+ {\bm \nabla} \cdot \left \{ \psi {\bm \nabla} \left [
\frac{2}{3} \tilde{B}_1 n \psi \right ] \right \} \nonumber\\
&& + {\bm \nabla} \cdot \left \{ (1+n)\psi {\bm \nabla} \left [
-(B_0^x-B_0^{\ell}+1) n + \left ( \tilde{B}_0 + \frac{2}{3} \tilde{B}_1
  \psi \right ) n^2 + B_0^x \left ( R_0^2\nabla^2 + 1 \right )^2 n
\right. \right. \nonumber\\
&& \left. \left. + B_0^x \left ( \alpha_2 R_0^2\nabla^2 + \frac{\alpha_4}{2}
  R_0^4\nabla^4 \right ) [(1+n)\psi] \right ] \right \}.
\label{eq:D2}
\end{eqnarray}
%%%%%%%%%%%%%%%%%
At this point in the derivation it should be noted that 
no additional approximations beyond those going into 
the expansions of Eq. (\ref{eq:C_expan}) have been introduced.

\subsubsection{Non-dimensional form of model}

To simplify the results, the above binary PFC equations can be
rescaled via defining a length scale $R_0$, a time scale
$R_0^2/M_1B_0^x$, $n \rightarrow \sqrt{v/B_0^x}~ n$, and $\psi
\rightarrow \sqrt{v/B_0^x}~ \psi$, yielding
\begin{equation}
\partial n / \partial t = {\cal D}_1 + m {\cal D}_2, \quad
\partial \psi / \partial t = \frac{1}{1 + g_0 n} \left [
(m - g_0 \psi) {\cal D}_1 + (1 - m g_0 \psi) {\cal D}_2 \right ],
\label{eq:bpfc_re}
\end{equation}
where
\begin{equation}
m=\frac{M_2}{M_1}=\frac{M_A-M_B}{M_A+M_B}, \quad
g_0 = \sqrt{\frac{B_0^x}{v}}, \quad 
v=\frac{2}{3} \tilde{B}_0
= \frac{\rho_l^2}{3} \hat{\bar{C}}_0^{(3)}.
\label{eq:m_g0_v}
\end{equation}
If keeping only terms up to 3rd order quantities of $n$ and
$\psi$, the results of ${\cal D}_1$ and ${\cal D}_2$ are rescaled as
\begin{eqnarray}
&{\cal D}_1 =& \nabla^2 \left \{ -\epsilon n + \left ( \nabla^2 +
    q_0^2 \right )^2 n + \left ( g_1 \psi + g \psi^2 \right ) n 
  + \left ( g_2 + \bar{g}_2 \psi \right ) n^2 + n^3 \right. \nonumber\\
&& \left. + \bar{g} \psi + v_1 \psi^2 + u_1 \psi^3 + \left
    (\alpha_2 \nabla^2 + \frac{\alpha_4}{2} \nabla^4 \right ) 
  \left [ (1 + g_0 n) \psi \right ] 
  + g_0 \psi \left (\alpha_2 \nabla^2 + \frac{\alpha_4}{2} \nabla^4
  \right ) n \right \} \nonumber\\ 
&& + g_0 {\bm \nabla} \cdot \left \{ n {\bm \nabla} \left [ 
\left ( \nabla^2 + q_0^2 \right )^2 n + \left (\alpha_2
    \nabla^2 + \frac{\alpha_4}{2} \nabla^4 \right ) \left (
    (1 + g_0 n) \psi \right ) \right ] \right \} \nonumber\\
&& + g_0 {\bm \nabla} \cdot \left [ \psi {\bm \nabla}
      \left ( - K_0 \nabla^2 + \kappa_0 \nabla^4 \right ) \left (
        (1 + g_0 n) \psi \right ) \right ]
- g_0 {\bm \nabla} \cdot \left [ \left ({\bm \nabla} \psi \right )
\left (\alpha_2 \nabla^2 + \frac{\alpha_4}{2} \nabla^4 \right ) n
\right ] \nonumber\\
&& + g_0^2 {\bm \nabla} \cdot \left \{ n \psi {\bm \nabla} \left [
    \left (\alpha_2 \nabla^2 + \frac{\alpha_4}{2} \nabla^4 \right )
      n + \left ( - K_0 \nabla^2 + \kappa_0 \nabla^4 \right ) \psi
      \right ] \right \},
\label{eq:re_D1}
\end{eqnarray}
\begin{eqnarray}
& {\cal D}_2 =& \nabla^2 \left \{ \bar{g} n + (1+ g_0 n)
    \left ( \alpha_2 \nabla^2 + \frac{\alpha_4}{2} \nabla^4 
    \right ) n + \left ( 2v_1 \psi + w_2 \psi^2 \right ) n
+ \left ( v_2 + g \psi \right ) n^2 + g_3 n^3
\right. \nonumber\\ 
&& \left. + w_0 \psi + v_0 \psi^2 + u_0 \psi^3 + \left (
 - K_0 \nabla^2 + \kappa_0 \nabla^4 \right ) \left [ (1 + g_0 n) \psi
\right ] \right \} \nonumber\\
&& + g_0 {\bm \nabla} \cdot \left \{ n {\bm \nabla} \left [
 w_0 \psi + \left ( - K_0 \nabla^2 + \kappa_0 \nabla^4 \right ) 
\left ( (1 + g_0 n) \psi \right ) \right ] \right \} 
- g_0 {\bm \nabla} \cdot \left [ \left ( {\bm \nabla} n \right )
\left (\alpha_2 \nabla^2 + \frac{\alpha_4}{2} \nabla^4 \right ) n
\right ] \nonumber\\
&& + g_0 {\bm \nabla} \cdot \left \{ \psi {\bm \nabla} \left [ 
\left ( -\epsilon + \gamma_1 \psi \right ) n + \gamma_2 n^2
+ \left ( \nabla^2 + q_0^2 \right )^2 n +
\left (\alpha_2 \nabla^2 + \frac{\alpha_4}{2} \nabla^4 \right )
\left ( (1 + g_0 n) \psi \right ) \right ] \right \} \nonumber\\
&& + g_0^2 {\bm \nabla} \cdot \left \{ n \psi {\bm \nabla} \left [
       \left ( \nabla^2 + q_0^2 \right )^2 n + \left
        (\alpha_2 \nabla^2 + \frac{\alpha_4}{2} \nabla^4 \right ) \psi
    \right ] \right \},
\label{eq:re_D2}
\end{eqnarray}
where the rescaled parameters are
\begin{eqnarray}
&& \epsilon = \frac{B_0^x - B_0^{\ell}}{B_0^x}, \qquad q_0 = 1, \quad
g = \frac{B_2^{\ell}}{v}, \quad
g_2=\frac{g_0}{2} \left ( \frac{2\tilde{B}_0-1}{B_0^x} - \epsilon 
\right ), \quad v_1=\left ( \frac{\beta_1+\Delta \beta}{2B_0^x}
\right ) g_0, \nonumber\\
&&w_0 = \frac{1+\beta_2}{B_0^x}, \quad
u_0 = \frac{2\tilde{B}_2}{9v}, \quad
K_0 = \frac{K}{B_0^x R_0^2} = -\frac{\Delta
  \hat{C}_2}{2\hat{\bar{C}}_2}, \quad
\kappa_0 = \frac{\kappa}{B_0^x R_0^4} 
= \frac{\Delta \hat{C}_4}{4\hat{\bar{C}}_4}, \nonumber\\
&& \bar{g}=\frac{\beta_0}{B_0^x}, \qquad
  g_1=\frac{B_1^{\ell}}{B_0^x} g_0, \qquad
\bar{g}_2=\frac{B_1^{\ell} + 2\tilde{B}_1}{2v}, \qquad
u_1 = \frac{B_3^{\ell}}{3v}, \nonumber\\
&& w_2 = \frac{\beta_0+2B_3^{\ell}}{v}, \qquad
v_2 = \left ( \frac{\beta_0/2+\tilde{B}_1/3}{B_0^x} \right ) g_0,
\qquad g_3 = \frac{2\tilde{B}_1}{9v}, \nonumber\\
&& v_0 = \left ( \frac{\beta_0+B_3^{\ell}}{2B_0^x} \right ) g_0, \qquad
\gamma_1 = 3 (g_3 - u_1) / g_0, \qquad
\gamma_2 = g_2 - v_1.
\label{eq:para_re}
\end{eqnarray}
Note that from Eqs. (\ref{eq:parameters}) and (\ref{eq:C_expan}),
$B_0^x$ can be rewritten as
\begin{equation}
B_0^x = \frac{\rho_l \hat{\bar{C}}_2^2}{4 \hat{\bar{C}}_4}
= \rho_l \left ( \hat{\bar{C}}_0 + \hat{\bar{C}}_{\rm max} \right ),
\label{eq:B0x}
\end{equation}
where $\hat{\bar{C}}_{\rm max}$ is the maximum of the first peak of
the two-point correlation function $\hat{\bar{C}}$ in Fourier
space. If $|\Delta \rho_l| = |\rho^A_l - \rho^B_l| \ll 
|\hat{\bar{C}}_0/\hat{\tilde{C}}_0^{(3)}|$, $B_0^{\ell} \sim 1 + \rho_l \hat{\bar{C}}_0$ 
from Eq. (\ref{eq:parameters}) and thus
\begin{equation}
\epsilon = \frac{B_0^x - B_0^{\ell}}{B_0^x} \sim \frac{\rho_l
  \hat{\bar{C}}_{\rm max} -1}{\rho_l \left ( \hat{\bar{C}}_0 +
    \hat{\bar{C}}_{\rm max} \right )} \sim
\frac{\hat{\bar{C}}_{\rm max}}{\hat{\bar{C}}_0 + \hat{\bar{C}}_{\rm
    max}}.
\end{equation}
Usually $\hat{\bar{C}}_{\rm max} \ll \hat{\bar{C}}_0$.
particularly when close to the melting point $T_m$, and hence $\epsilon$ can be
viewed as a small variable (also used in amplitude equation expansion
given below), proportional to $(T-T_m)/T_m$ as discussed in the
original PFC model \cite{re:elder07}.  

\subsubsection{Simplification of scaled binary model}
\label{sec:simplify}

The rescaled PFC dynamic equations (\ref{eq:bpfc_re})--(\ref{eq:re_D2})
can be further simplified to a lower order form via a scale analysis. 
A simple scale analysis of Eqs. (\ref{eq:re_D1}) and
(\ref{eq:re_D2}) yields $n, \psi \sim {\cal O}(\epsilon^{1/2})$ (e.g.,
from Eq. (\ref{eq:re_D1}) we have ${\cal O}(\epsilon n) \sim 
{\cal O}(n^3)$ and ${\cal O}(\psi) \sim {\cal O}(n)$, as is usually
assumed). To simplify the results the following approximations 
are made: (i) Assume that $\{ |\hat{\bar{C}}_0|, |\hat{\bar{C}}_0^{(3)}|, 
|\delta \hat{C}_0^{(3)}| \} \gg \{ |\delta \hat{C}_0|,
|\hat{\tilde{C}}_0^{(3)}|, |\Delta \hat{C}_0^{(3)}| \}$ and 
$|\rho_l^A - \rho_l^B|\ll|\rho_l^A+\rho_l^B|$.
(An example case would be that the zeroth-mode ($q=0$) correlation functions 
between the same atomic species are of the same order, and are either 
significantly larger or significantly smaller than those between different ones; see
Eq. (\ref{eq:CAB}).) Thus for the rescaled parameters 
in Eq. (\ref{eq:para_re}), we can estimate (based on the
definitions in Eqs. (\ref{eq:parameters}) and (\ref{eq:CAB}), as well
as Eqs. (\ref{eq:m_g0_v}) and (\ref{eq:B0x})) that
$$
g_0, g, g_2, v_1, u_0 \sim {\cal O}(1) ~ {\rm or} ~ {\cal
  O}(\epsilon^{1/2}), \qquad
\bar{g}, g_1, \bar{g}_2, u_1, w_2, v_2, g_3, v_0, \gamma_1,
\gamma_2 \sim {\cal O}(\epsilon) ~{\rm or ~ higher}.
$$
(ii) The concentration field $\psi$ is slowly varying in space, and we
can keep only the lowest linear gradient terms for $\psi$. (iii) Similar to
the single-component case in Sec. \ref{sec:ddft_pure}, it can be
argued that in Eqs. (\ref{eq:re_D1}) and (\ref{eq:re_D2}), compared to
the first terms $\nabla^2 \{ \cdots \}$, all other terms ($g_0 {\bm
  \nabla} \cdot \{ \cdots \}$) are of higher order. (iv) For
linear terms in $n$, only $[-\epsilon + (\nabla^2 + q_0^2)^2] n$ is
kept which will lead to periodic crystal structure in solid phases,
while the other term $(\alpha_2 \nabla^2 + \frac{\alpha_4}{2}
\nabla^4) n$ is neglected, which corresponds to ignoring  the
$n\psi$ related terms in the free energy functional (\ref{eq:F_ndN}) 
owing to to the much larger length scales of $\psi$ field \cite{re:elder07}. 
(v) It is assumed that $\alpha_2 \simeq \alpha_4/2 \simeq 2
\alpha$ (see the discussions below Eq. (\ref{eq:alpha})).  

To lowest order in ${\cal O}(\epsilon^{3/2})$  the above
simplifications reduce the PFC equations (\ref{eq:bpfc_re}),
(\ref{eq:re_D1}), and (\ref{eq:re_D2}) to
\begin{equation}
\partial n / \partial t = {\cal D}_1 + m {\cal D}_2, \quad
\partial \psi / \partial t = m {\cal D}_1 + {\cal D}_2,
\label{eq:pfc_npsi}
\end{equation}
where
\begin{eqnarray}
&{\cal D}_1 = \nabla^2 \left \{ \left ( - \epsilon + g \psi^2 \right )
  n + \left ( \nabla^2 + q_0^2 \right )^2 n + g_2 n^2 + n^3 + v_1 \psi^2
  + \etaa \left [ \psi \left ( \nabla^2 + \nabla^4 \right ) n + \left (
    \nabla^2 + \nabla^4 \right ) (n \psi) \right ] \right \}, & \nonumber\\
&{\cal D}_2 = \nabla^2 \left [ \etaa n \left ( \nabla^2 + \nabla^4
  \right ) n + \left ( w_0 + 2v_1 n + g n^2 \right ) \psi + u_0
  \psi^3 - K_0 \nabla^2 \psi \right ],& \label{eq:D12}
\end{eqnarray}
with $\alpha_0 = g_0 \alpha$ the rescaled solute expansion coefficient.
Equations (\ref{eq:pfc_npsi}) and (\ref{eq:D12}) recover 
the original binary PFC model with conserved dynamics for
both $n$ and $\psi$ fields \cite{re:elder07,re:elder10}, except for
the $v_1$ terms ($v_1 \psi^2$ and $2v_1 n \psi$), which account
for additional coupling between the atomic density and concentration
fields (or between small crystalline and ``slow'' concentration
scales). This can also be seen via rewriting Eq. (\ref{eq:D12})
through an effective potential (or free energy functional)
${\cal F}_{\rm eff}$:
\begin{eqnarray}
&{\cal D}_1 = \nabla^2 \frac{\delta {\cal F}_{\rm eff}}{\delta n},
\qquad {\cal D}_2 = \nabla^2 \frac{\delta {\cal F}_{\rm eff}}{\delta \psi},& 
\label{eq:D12_eff} \\
&{\cal F}_{\rm eff} = \int d{\bm r} \left \{ - \frac{1}{2} \epsilon
n^2 + \frac{1}{2} n \left ( \nabla^2 + q_0^2 \right )^2 n
+ \frac{1}{3} g_2 n^3 + \frac{1}{4} n^4
+ \etaa n \left ( \nabla^2 + \nabla^4 \right ) (n \psi)
\right.& \nonumber\\
&\left. + \frac{1}{2} (w_0 + 2v_1 n + g n^2) \psi^2 
+ \frac{1}{4} u_0 \psi^4
+ \frac{1}{2} K_0 \left | {\bm \nabla} \psi \right |^2 \right \}.&
\label{eq:F_eff}
\end{eqnarray}
In the rest of this work, all results, including the corresponding
amplitude equation formalism, noise dynamics, and the related
applications, are based on the simplified PFC dynamic equations
(\ref{eq:pfc_npsi}) and (\ref{eq:D12}).

The above results can also be derived and verified through two other
alternative methods, as given in Appendix
\ref{append:altern}. Furthermore, to include
higher-order terms (e.g., $n^4$, $\psi^4$, ...) in both the free
energy functional and the dynamic equations, we need to consider 
higher-order direct correlation functions (4-point, 5-point, etc.) as
shown in the single-component case (Sec. \ref{sec:ddft_pure}), with
similar derivation steps.

\section{Amplitude equation formalism for binary PFC model}
\label{sec:ampl}

As discussed in Sec. \ref{sec:intro}, the PFC methodology includes
model equations governing the dynamics of density and concentration fields as 
given above. This section will examine the long wavelength and time limits of the 
alloy PFC  model by deriving its corresponding amplitude equations, which 
emerge after coarse-graining the model using a multiple-scale analysis. 
The amplitude representation for single-component PFC models has
been well established \cite{re:goldenfeld05,re:athreya06,re:athreya07,re:chan10,re:yeon10},
while for binary systems the corresponding amplitude equations have
been derived very recently, for both 2D hexagonal/triangular and 3D
bcc and fcc crystalline structures \cite{re:elder10,re:spatschek10}.
Here we focus on the 2D amplitude equations for the binary PFC model
with hexagonal lattice structure, yielding a complete formulation
incorporating the effects of different mobilities between alloy components 
and dynamic variation of the average atomic density, both of which are missing 
in the previous binary alloy amplitude formulation \cite{re:elder10}.  
It is straightforward to extend this calculation to 3D bcc or fcc 
structures. The derivation process involves two steps: the standard multiple
scale expansion \cite{re:cross93} for lowest order amplitude equations
(Sec. \ref{sec:multiscale}), and a new hybrid approach (combining
results of multiple scale approach and the idea of ``Quick and Dirty''
renormalization-group (RG) method developed by Goldenfeld \textit{et al.}
\cite{re:goldenfeld05,re:athreya06}) for full order amplitude
equations (see Sec. \ref{sec:hybrid}). To apply the multiple 
scale analysis, the rescaled PFC equations (\ref{eq:pfc_npsi}) and
(\ref{eq:D12}) are used.

\subsection{Multiple scale expansion: Lowest order amplitude equations}
\label{sec:multiscale}

Following the standard procedure of multiple scale approach
\cite{re:cross93}, in the 
limit of small $\epsilon$ (i.e., high temperature) we can separate
``slow'' spatial and temporal scales $(X=\epsilon^{1/2}x,
Y=\epsilon^{1/2}y, T=\epsilon t)$ for structural profile/amplitudes
from ``fast'' scales of the underlying crystalline
lattice. Substituting
\begin{equation}
\partial_x \rightarrow \partial_x + \epsilon^{1/2} \partial_X, \qquad
\partial_y \rightarrow \partial_y + \epsilon^{1/2} \partial_Y, \qquad
\partial_t \rightarrow \epsilon \partial_T, \label{eq:deriv_xyt}
\end{equation}
and the expansions
\begin{eqnarray}
& n = \epsilon^{1/2} n^{(1/2)} + \epsilon n^{(1)} + \epsilon^{3/2}
n^{(3/2)} + \epsilon^2  n^{(2)} + \cdots, & \nonumber\\
& \psi = \epsilon^{1/2} \psi^{(1/2)} + \epsilon \psi^{(1)} +
\epsilon^{3/2} \psi^{(3/2)} + \epsilon^2  \psi^{(2)} + \cdots, &
\end{eqnarray}
into the PFC equations (\ref{eq:pfc_npsi}) and (\ref{eq:D12}), we can
obtain the corresponding equations at each order of
$\epsilon^{1/2}$. For simplicity, assume $m, \alpha_0, g, u_0, K_0
\sim {\cal O}(1)$, $g_2, v_1 \sim {\cal O}(\epsilon^{1/2})$, and $w_0
\sim {\cal O}(\epsilon)$ (as also assumed in Sec. \ref{sec:simplify}
for model simplification).
To ${\cal O}(\epsilon^{1/2})$ and ${\cal O}(\epsilon)$ we have
\begin{equation}
\nabla^2 \left [ {\cal L}_0 n^{(i)} - m K_0 \nabla^2
  \psi^{(i)} \right ] =0, \qquad
\nabla^2 \left [ m {\cal L}_0 n^{(i)} - K_0 \nabla^2
  \psi^{(i)} \right ] =0,
\end{equation}
where $i=1/2$ or $1$, and ${\cal L}_0 = (\nabla^2 +q_0^2)^2$. This
leads to $(1-m^2) \nabla^2 {\cal L}_0 n^{(i)} = 0$ and $(1-m^2)
\nabla^4 \psi^{(i)} = 0$, with solutions
\begin{equation}
n^{(i)} = n_0^{(i)}(X,Y,T) + \sum_{j=1}^{3} \A_j^{(i)}(X,Y,T) e^{i
  {\bm q}_j^0 \cdot {\bm r}} + {\rm c.c.}, \qquad
\psi^{(i)} = \psi_0^{(i)}(X,Y,T), \label{eq:npsi_1}
\end{equation}
where ${\bm q}_j^0$ represent the three reciprocal lattice 
vectors for 2D hexagonal/triangular structure:
${\bm q_1^0} = -q_0 ( \sqrt{3} \hat{x}/2 + \hat{y}/2 )$, ${\bm q_2^0} =
q_0 \hat{y}$, ${\bm q_3^0} = q_0 (\sqrt{3} \hat{x}/2 - \hat{y}/2 )$. 
$A_j$ are the slowly varying complex amplitudes of the modes 
${\bm q}_j^0$, while $n_0$ and $\psi_0$ refer to the real amplitudes of
the zero wavenumber neutral mode as a result of order parameter
conservation \cite{re:matthews00}.

Expanding to ${\cal O}(\epsilon^{3/2})$ yields (with
${\bm \nabla}_s = (\partial_X, \partial_Y)$, ${\bm \nabla} \cdot {\bm
  \nabla}_s = \partial_x \partial_X + \partial_y \partial_Y$, and
$\nabla_s^2 = \partial_X^2 + \partial_Y^2$)
\begin{eqnarray}
\nabla^2 {\cal L}_0 n^{(3/2)} - m K_0 \nabla^4 \psi^{(3/2)}
&=& \partial_T n^{(1/2)} + \left [ \nabla^2 - \nabla^2
  \left ( 2 {\bm \nabla} \cdot {\bm \nabla}_s \right )^2 - q_0^4
  \nabla_s^2 \right ] n^{(1/2)} - \nabla^2 \left [ g_2 {n^{(1/2)}}^2
  + {n^{(1/2)}}^3 \right ] \nonumber\\
&-& g {\psi^{(1/2)}}^2 \nabla^2 n^{(1/2)}
+ \etaa \nabla^2 \left [ \psi^{(1/2)} \left ( 2 {\bm \nabla} \cdot
    {\bm \nabla}_s \right ) n^{(1/2)} + \left ( 2 {\bm \nabla} \cdot
    {\bm \nabla}_s \right ) \left ( \psi^{(1/2)} n^{(1/2)} \right ) 
\right ] \nonumber\\
&+& m \left \{ \etaa \nabla^2 \left [ n^{(1/2)} 
\left ( 2 {\bm \nabla} \cdot {\bm \nabla}_s \right ) n^{(1/2)} \right ]
- g \psi^{(1/2)} \nabla^2 {n^{(1/2)}}^2 
- 2v_1 \psi^{(1/2)} \nabla^2 n^{(1/2)} \right \}, \nonumber\\
m \nabla^2 {\cal L}_0 n^{(3/2)} - K_0 \nabla^4 \psi^{(3/2)}
&=& m \left \{ \left [ \nabla^2 - \nabla^2
  \left ( 2 {\bm \nabla} \cdot {\bm \nabla}_s \right )^2 - q_0^4
  \nabla_s^2 \right ] n^{(1/2)} - \nabla^2 \left [ g_2 {n^{(1/2)}}^2
  + {n^{(1/2)}}^3 \right ] \right. \nonumber\\
&-& \left. g {\psi^{(1/2)}}^2 \nabla^2 n^{(1/2)}
+ \etaa \nabla^2 \left [ \psi^{(1/2)} \left ( 2 {\bm \nabla} \cdot
    {\bm \nabla}_s \right ) n^{(1/2)} + \left ( 2 {\bm \nabla} \cdot
    {\bm \nabla}_s \right ) \left ( \psi^{(1/2)} n^{(1/2)} \right ) 
\right ] \right \} \nonumber\\
&+& \partial_T \psi^{(1/2)} + \etaa \nabla^2 \left [ n^{(1/2)} 
\left ( 2 {\bm \nabla} \cdot {\bm \nabla}_s \right ) n^{(1/2)} \right ]
- g \psi^{(1/2)} \nabla^2 {n^{(1/2)}}^2
- 2v_1 \psi^{(1/2)} \nabla^2 n^{(1/2)}, \nonumber
\end{eqnarray}
which is equivalent to
\begin{eqnarray}
(1-m^2) \nabla^2 {\cal L}_0 n^{(3/2)} &=& \partial_T n^{(1/2)}
- m \partial_T \psi^{(1/2)} + (1-m^2) \left \{ \left [ \nabla^2 - \nabla^2
  \left ( 2 {\bm \nabla} \cdot {\bm \nabla}_s \right )^2 - q_0^4
  \nabla_s^2 \right ] n^{(1/2)} - g {\psi^{(1/2)}}^2 \nabla^2
  n^{(1/2)} \right. \nonumber\\
&-& \left. \nabla^2 \left [ g_2 {n^{(1/2)}}^2 + {n^{(1/2)}}^3 \right ]
+ \etaa \nabla^2 \left [ \psi^{(1/2)} \left ( 2 {\bm \nabla} \cdot
    {\bm \nabla}_s \right ) n^{(1/2)} + \left ( 2 {\bm \nabla} \cdot
    {\bm \nabla}_s \right ) \left ( \psi^{(1/2)} n^{(1/2)} \right ) 
\right ] \right \}, \nonumber\\
(1-m^2) K_0 \nabla^4 \psi^{(3/2)} &=& m \partial_T n^{(1/2)}
- \partial_T \psi^{(1/2)} \nonumber\\
&+& (1-m^2) \left \{ - \etaa \nabla^2 \left [
    n^{(1/2)} \left ( 2 {\bm \nabla} \cdot {\bm \nabla}_s \right )
    n^{(1/2)} \right ] + g \psi^{(1/2)} \nabla^2 {n^{(1/2)}}^2 
    + 2v_1 \psi^{(1/2)} \nabla^2 n^{(1/2)} \right \}.
\label{eq:expan32}
\end{eqnarray}
As shown in Eq. (\ref{eq:npsi_1}), the zero eigenvectors of operators
$\nabla^2 {\cal L}_0$ and $\nabla^4$ are $(e^{\pm i {\bm q}_j^0 \cdot
  {\bm r}}, 1)$ and $1$ (of the 0th mode), respectively. 
Using the Fredholm theory or solvability condition \cite{re:cross93} in
Eq. (\ref{eq:expan32}), we can derive the lowest order amplitude
equations as (with $j=1,2,3$)
\begin{eqnarray}
& \partial A_j^{(1/2)} / \partial t = - (1-m^2) q_0^2 \left \{ 
\left [ -1 + \left ( 2i {\bm q}_j^0 \cdot {\bm \nabla}_s \right )^2
\right ] A_j^{(1/2)} + \left [ 3 {n_0^{(1/2)}}^2 + 2g_2 n_0^{(1/2)}
+ g {\psi_0^{(1/2)}}^2 \right ] A_j^{(1/2)} \right. & \nonumber\\
& + 3 A_j^{(1/2)} \left [ \left | A_j^{(1/2)} \right |^2 + 
2 \sum_{k,l \neq j}^{k<l} \left ( \left | A_k^{(1/2)} \right |^2 
+ \left | A_l^{(1/2)} \right |^2 \right ) \right ] + 
\left ( 6 n_0^{(1/2)} + 2g_2 \right )
\sum_{k,l \neq j}^{k<l} {A_k^{(1/2)}}^* {A_l^{(1/2)}}^* & \nonumber\\
& \left. - \etaa \left [ \psi_0^{(1/2)} \left ( 2i {\bm q}_j^0 \cdot 
{\bm \nabla}_s \right ) A_j^{(1/2)} + \left ( 2i {\bm q}_j^0 \cdot 
{\bm \nabla}_s \right ) \left ( \psi_0^{(1/2)} A_j^{(1/2)} \right )
\right ] \right \}, & \label{eq:Aj_} \\
& \partial n_0^{(1/2)} / \partial t = q_0^4 \nabla_s^2 n_0^{(1/2)},
& \label{eq:n0_} \\ 
& \partial \psi_0^{(1/2)} / \partial t = m q_0^4 \nabla_s^2
\psi_0^{(1/2)}. & \label{eq:psi0_} 
\end{eqnarray}
Using the scaling relation $A_j = \epsilon^{1/2} A_j^{(1/2)}$,
$n_0 = \epsilon^{1/2} n_0^{(1/2)}$, and $\psi_0 = \epsilon^{1/2}
\psi_0^{(1/2)}$, we can then obtain the corresponding amplitude
equations in the unscaled units $(x,y,t)$.

It is noted that the direct solutions to Eq. (\ref{eq:expan32}) have the form
\begin{eqnarray}
& n^{(3/2)} = n_0^{(3/2)}(X,Y,T) + \sum\limits_{j=1}^{3}
\A_j^{(3/2)}(X,Y,T) e^{i {\bm q}_j^0 \cdot {\bm r}} + {\rm c.c.} 
+ {\rm higher~ harmonics},& \label{eq:n_32}\\
& \psi^{(3/2)} = \psi_0^{(3/2)}(X,Y,T) + \sum\limits_{j=1}^{3}
\psi_j^{(3/2)}(X,Y,T) e^{i {\bm q}_j^0 \cdot {\bm r}} + {\rm c.c.} 
+ {\rm higher~ harmonics}.& \label{eq:psi_32}
\end{eqnarray}
Compared to Eq. (\ref{eq:npsi_1}) for the ${\cal O}(\epsilon^{1/2})$
and ${\cal O}(\epsilon)$ solutions, it can be found that the complex
amplitudes $\psi_j$ corresponding to the periodic modes of the concentration 
field in substitutional binary alloys considered here is generally of 
order $\epsilon$ higher than $A_j$, $n_0$, and $\psi_0$.  
For systems in which a sublattice ordering occurs (such as 
B2 or B32 ordering in bcc crystals), $\psi_j$  would 
be of the same order as these other fields.  To describe sublattice 
ordering a different free energy functional from the one 
given in Eq. (\ref{eq:F_eff}) would also be required. Detailed
results will be presented elsewhere.

\subsection{A hybrid approach: Full order amplitude equations}
\label{sec:hybrid}

The lowest order amplitude equations (\ref{eq:Aj_})--(\ref{eq:psi0_})
derived above are not sufficient to describe the evolution of binary
systems; e.g., Eq. (\ref{eq:Aj_}) for $A_j$ is not rotationally
invariant, and Eqs. (\ref{eq:n0_}) and (\ref{eq:psi0_}) for $n_0$ and
$\psi_0$ are just diffusion equations and would lead to a
steady state solution of constant $n_0$ and $\psi_0$ values
at long enough time. We thus need higher order
amplitude equations, which in principle can be derived by extending
the multiple scale process described above to higher order
expansions. However, the procedure is complicated and tedious.
In the following we use, instead, a simplified approach combining the above
steps of multiple scale expansion and the idea of the ``Quick and Dirty''
RG method \cite{re:goldenfeld05,re:athreya06}.

The first step is the standard multiple scale expansion given in
Sec. \ref{sec:multiscale}, starting from the scale separation
Eq. (\ref{eq:deriv_xyt}). From the solution forms of 
Eqs. (\ref{eq:npsi_1}), (\ref{eq:n_32}) and (\ref{eq:psi_32}), we know
that to all orders of $\epsilon$ the solutions of $n$ and $\psi$
fields can be written as
\begin{eqnarray}
& n = n_0(X,Y,T) + \sum\limits_{j=1}^{3} \A_j(X,Y,T)
e^{i {\bm q}_j^0 \cdot {\bm r}} + {\rm c.c.} 
+ {\rm higher~ harmonics},& \label{eq:n_expan}\\
& \psi = \psi_0(X,Y,T) + \sum\limits_{j=1}^{3}
\psi_j(X,Y,T) e^{i {\bm q}_j^0 \cdot {\bm r}} + {\rm c.c.} 
+ {\rm higher~ harmonics},& \label{eq:psi_expan}
\end{eqnarray}
with $(X,Y,T)$ the slow scales. Thus, based on the separation between
``fast''/''slow'' scales the following expansions (full-order) can be
obtained:
\begin{eqnarray}
\nabla^2 n &\rightarrow& \epsilon \nabla_s^2 n_0 + \sum_{j=1}^{3}
({\cal L}_j^s \A_j) e^{i {\bm q}_j^0 \cdot {\bm r}} + {\rm c.c.}
+ \{\cdots\}, \nonumber\\
(\nabla^2+q_0^2)^2 n &\rightarrow& \left ( \epsilon \nabla_s^2 + q_0^2
  \right )^2 n_0 + \sum_{j=1}^{3} ({{\cal G}_j^s}^2 \A_j)
e^{i {\bm q}_j^0 \cdot {\bm r}} + {\rm c.c.} + \{\cdots\}, 
\nonumber\\
(\nabla^2 + \nabla^4) (n \psi) &\rightarrow& \left ( 
\epsilon \nabla_s^2 + \epsilon^2 \nabla_s^4 \right ) \left (
n_0 \psi_0 + \sum_{j=1}^3 \A_j \psi_j^* + {\rm c.c.} \right )
\nonumber\\
&+& \sum_{j=1}^3 \left[ {\cal L}_j^s {\cal G}_j^s \left ( \psi_0 \A_j + n_0
  \psi_j + \sum_{k,l \neq j}^{k<l} \A_k^* \psi_l^* \right )\right]
e^{i {\bm q}_j^0 \cdot {\bm r}} + {\rm c.c.} + \{\cdots\}, \nonumber\\
n^2 &\rightarrow& n_0^2 + 2 \sum_{j=1}^3 |\A_j|^2 + \sum_{j=1}^3 \left
  ( 2n_0 \A_j + 2 \sum_{k,l \neq j}^{k<l} \A_k^* \A_l^* \right )
e^{i {\bm q}_j^0 \cdot {\bm r}} + {\rm c.c.} + \{\cdots\}, \nonumber\\
n^3 &\rightarrow& n_0^3 + 6n_0 \sum_{j=1}^3 |\A_j|^2 + 6 \left (
  \prod_{j=1}^3 \A_j + {\rm c.c.} \right ) \nonumber\\
&+& \sum_{j=1}^3 \left \{ 3(n_0^2 + |\A_j|^2) \A_j + 6 \sum_{k,l \neq
    j}^{k<l} \left [ n_0 \A_k^* \A_l^* + \A_j \left ( |\A_k|^2 + |\A_l|^2
    \right ) \right ] \right \} 
e^{i {\bm q}_j^0 \cdot {\bm r}} + {\rm c.c.} + \{\cdots\}, \nonumber\\
n \psi^2 &\rightarrow& n_0 \psi_0^2 + 2 n_0 \sum_{j=1}^3 |\psi_j|^2
+ \sum_{j=1}^3 \left ( 2 \psi_0 \A_j + \sum_{k \neq l \neq j} \A_k^*
  \psi_l^* \right ) \psi_j^* + {\rm c.c.}
+ \sum_{j=1}^3 \left [ 2n_0 \left ( \psi_0 \psi_j +
    \sum_{k,l \neq j}^{k<l} \psi_k^* \psi_l^* \right )
\right. \nonumber\\
&+& \left. \left ( \psi_0^2 + 2 \sum_{k=1}^3
    |\psi_k|^2 \right ) \A_j
  + 2 \psi_0 \sum_{k \neq l \neq j} \A_k^* \psi_l^* + 2 \psi_j 
  \sum_{k \neq j} \left ( \A_k \psi_k^* + {\rm c.c.} \right )
  + \A_j^* \psi_j^2 \right ]
e^{i {\bm q}_j^0 \cdot {\bm r}} + {\rm c.c.} + \{\cdots\}, \nonumber\\
\cdots \cdots && \label{eq:expan}
\end{eqnarray}
where $\{\cdots\}$ refers to the contributions from higher harmonics
and the slow operators are given by  
\begin{equation}
{\cal L}_j^s = \epsilon \nabla_s^2 + \epsilon^{1/2}
\left ( 2i {\bm q}_j^0 \cdot {\bm \nabla}_s \right ) - q_0^2, \qquad
{\cal G}_j^s = {\cal L}_j^s + q_0^2 = \epsilon \nabla_s^2 +
\epsilon^{1/2} \left ( 2i {\bm q}_j^0 \cdot {\bm \nabla}_s \right ).
\label{eq:LGs}
\end{equation}

Assuming that higher harmonic terms can be neglected, the binary PFC
equations (\ref{eq:pfc_npsi}) and (\ref{eq:D12}) are then replaced by
\begin{eqnarray}
& \epsilon \partial_T n_0 + \epsilon \sum_j \partial_T \A_j
e^{i {\bm q}_j^0 \cdot {\bm r}} + {\rm c.c.} = {\cal D}_1^s 
+ m {\cal D}_2^s,& \label{eq:n_s}\\
& \epsilon \partial_T \psi_0 + \epsilon \sum_j \partial_T \psi_j
e^{i {\bm q}_j^0 \cdot {\bm r}} + {\rm c.c.} = m {\cal D}_1^s 
+ {\cal D}_2^s,& \label{eq:psi_s}
\end{eqnarray}
where ${\cal D}_1^s$ and ${\cal D}_2^s$ are the corresponding
expansion of ${\cal D}_1$ and ${\cal D}_2$, as obtained by
substituting Eq. (\ref{eq:expan}) into Eq. (\ref{eq:D12}).
Integrating Eqs.~(\ref{eq:n_s}) and (\ref{eq:psi_s}) over the eigenmodes 
$\int d{\bm r} \{ e^{-i {\bm q}_j^0 \cdot {\bm r}}, 1 \}$, keeping in
mind that ``fast'' and ``slow'' scales are separated, and in the final
step returning to original unscaled units $(x,y,t)$, we arrive at the 
following full-order amplitude equations for the binary PFC model:
\begin{eqnarray}
\partial n_0 / \partial t &=& \nabla^2 \frac{\delta {\cal F}}
{\delta n_0} + m \nabla^2 \frac{\delta {\cal F}}{\delta \psi_0}, 
\label{eq:n0}\\
\partial \A_j / \partial t &=& {\cal L}_j \left ( 
\frac{\delta {\cal F}}{\delta \A_j^*} + m \frac{\delta {\cal F}}
{\delta \psi_j^*} \right ) \simeq - q_0^2 \left ( 
\frac{\delta {\cal F}}{\delta \A_j^*} + m \frac{\delta {\cal F}}
{\delta \psi_j^*} \right ), \label{eq:Aj}\\
\partial \psi_0 / \partial t &=& m \nabla^2 \frac{\delta {\cal F}}
{\delta n_0} + \nabla^2 \frac{\delta {\cal F}}{\delta \psi_0}, 
\label{eq:psi0}\\
\partial \psi_j / \partial t &=& {\cal L}_j \left ( 
m \frac{\delta {\cal F}}{\delta \A_j^*} + \frac{\delta {\cal F}}
{\delta \psi_j^*} \right ) \simeq - q_0^2 \left ( 
m \frac{\delta {\cal F}}{\delta \A_j^*} + \frac{\delta {\cal F}}
{\delta \psi_j^*} \right ), \label{eq:psij}
\end{eqnarray}
where $j=1,2,3$, and
\begin{eqnarray}
& {\cal F} = \int d{\bm r} & \left \{ - \frac{1}{2} \epsilon n_0^2 +
  \frac{1}{2} \left [ \left ( \nabla^2 + q_0^2 \right ) n_0 \right ]^2
  + \frac{1}{3} g_2 n_0^3 + \frac{1}{4} n_0^4
  + \left ( - \epsilon + 3 n_0^2 + 2 g_2 n_0 + g \psi_0^2 \right )
  \sum_{j=1}^3 |\A_j|^2 \right. \nonumber\\
&& + \sum_{j=1}^3 \left | {\cal G}_j \A_j \right |^2 + \frac{3}{2}
\sum_{j=1}^3 |\A_j|^4 + (6n_0 + 2g_2) \left ( \prod_{j=1}^3 \A_j +
  {\rm c.c.} \right ) + 6 \sum_{j<k} |\A_j|^2 |\A_k|^2 \nonumber\\
&& + g \left [ \frac{1}{2} n_0^2 \psi_0^2 + n_0^2 \sum_{j=1}^3
  |\psi_j|^2 + 2 \sum_{j,k=1}^3 |\A_j|^2 |\psi_k|^2 + \sum_{j=1}^3
  \left ( 2n_0 \psi_0 \A_j \psi_j^* + \frac{1}{2} \A_j^2
    {\psi_j^*}^2 + {\rm c.c.} \right ) \right. \nonumber\\
&& \left. \qquad + \sum_{j \neq k} (\A_j \psi_j^* + {\rm c.c.})
(\A_k \psi_k^* + {\rm c.c.}) + \sum_{j \neq k \neq l} ( n_0 \psi_j^*
+ \psi_0 \A_j^* ) \A_k^* \psi_l^* + {\rm c.c.} \right ] \nonumber\\
&& + \etaa \left [ \psi_0 n_0 \left ( \nabla^2 + \nabla^4 \right )
  n_0 + \psi_0 \left ( \sum_{j=1}^3 \A_j^* {\cal L}_j {\cal G}_j \A_j
    + {\rm c.c.} \right ) + n_0 \left ( \nabla^2 + \nabla^4 \right )
  \left ( \sum_{j=1}^3 \A_j \psi_j^* + {\rm c.c.} \right ) \right.
  \nonumber\\
&& \left. \qquad \quad + n_0 \sum_{j=1}^3 \psi_j^* {\cal L}_j {\cal G}_j
  \A_j + \sum_{j \neq k \neq l} \A_j \psi_k {\cal L}_l {\cal G}_l
  \A_l + {\rm c.c.} \right ] \nonumber\\
&& + \frac{1}{2} w_0 \psi_0^2 + \frac{1}{2} K_0 \left | \nabla
  \psi_0 \right |^2 + \frac{1}{4} u_0 \psi_0^4 
+ \left ( w_0 + 3 u_0 \psi_0^2 \right ) \sum_{j=1}^3
|\psi_j|^2 - \frac{1}{2} K_0 \sum_{j=1}^3 \left ( \psi_j {\cal
    L}_j^* \psi_j^* + {\rm c.c.} \right ) \nonumber\\
&& + u_0 \left [ \frac{3}{2} \sum_{j=1}^3 |\psi_j|^4 + 6 \psi_0
  \left ( \prod_{j=1}^3 \psi_j + {\rm c.c.} \right )
  + 6 \sum_{j<k} |\psi_j|^2 |\psi_k|^2 \right ] \nonumber\\
&& \left. + v_1 \left [ n_0 \psi_0^2 + 2 n_0 \sum_{j=1}^3 |\psi_j|^2
   + 2 \psi_0 \left ( \sum_{j=1}^3 A_j \psi_j^* + {\rm c.c.} \right )
   + \sum_{j \neq k \neq l} A_j \psi_k \psi_l + {\rm c.c.} \right ]
\right \}. \label{eq:F}
\end{eqnarray}
Corresponding to Eq. (\ref{eq:LGs}), the operators ${\cal L}_j$ and
${\cal G}_j$ (in the original scales) are defined by
\begin{equation}
{\cal L}_j = \nabla^2 + 2i {\bm q}_j^0 \cdot {\bm \nabla} - q_0^2,
\qquad {\cal G}_j = {\cal L}_j + q_0^2 = \nabla^2 +
2i {\bm q}_j^0 \cdot {\bm \nabla}, \label{eq:L_G}
\end{equation}
and for simplicity, in Eqs. (\ref{eq:Aj})--(\ref{eq:F}) the operator
${\cal L}_j$ can be replaced by $-q_0^2$ in the long wavelength
approximation as adopted in Ref. \cite{re:elder10}.

As discussed at the end of Sec. \ref{sec:multiscale}, the amplitudes
$\psi_j$ are of ${\cal O}(\epsilon)$ higher compared to the
others for the free energy functional considered here.
Thus the above amplitude equations can be further simplified
by assuming $\psi_j \sim 0$, which leads to
\begin{eqnarray}
\partial \A_j / \partial t &=& - q_0^2 \frac{\delta {\cal F}}{\delta
  \A_j^*} - m q_0^2 \left \{ \etaa \left [ \A_j \left ( \nabla^2 +
      \nabla^4 \right ) n_0 + n_0 {\cal L}_j {\cal G}_j \A_j
    + \sum_{k \neq l \neq j} \A_k^* {\cal L}_l^* {\cal G}_l^* \A_l^*
  \right ] \right. \nonumber\\
&+& \left. 2g \psi_0 ( n_0 \A_j + \sum_{k,l \neq j}^{k<l}
    \A_k^* \A_l^* ) + 2v_1 \psi_0 A_j \right \} \nonumber\\
&=& - q_0^2 \frac{\delta {\cal F}}{\delta \A_j^*} - m q_0^2 
\left. \frac{\delta {\cal F}}{\delta \psi_j^*} \right |_{\psi_j=0}.
\nonumber
\end{eqnarray}
The dynamic equations for $n_0$ and $\psi_0$ are still governed by
Eqs. (\ref{eq:n0}) and (\ref{eq:psi0}). 
The amplitude equations can be further simplified by noting from
Eq. (\ref{eq:psij})
$0 \simeq \partial \psi_j / \partial t = -q_0^2 ( m \delta {\cal F} /
\delta \A_j^* + \delta {\cal F} / \delta \psi_j^* |_{\psi_j=0})$.
Thus, the above dynamic equation for $A_j$ can be further approximated as
\begin{equation}
\partial \A_j / \partial t \simeq - q_0^2 (1-m^2) 
\frac{\delta {\cal F}}{\delta \A_j^*},
\label{eq:Aj0}
\end{equation}
which to lowest order recovers the result of multiple scale
approach given in Eq. (\ref{eq:Aj_}). In the applications that will be
examined in Sec. \ref{sec:appl} the simplified amplitude equations
(\ref{eq:n0}), (\ref{eq:psi0}), and (\ref{eq:Aj0}) will be used.

\section{Noise dynamics and stochastic amplitude equations}
\label{sec:noise}

In the original PFC model \cite{re:elder02,re:elder04} a conserved
noise dynamics has been incorporated. However, in DDFT it has been
argued that the dynamic equation governing the density field evolution
should be deterministic and an additional stochastic noise term added
to Eq. (\ref{eq:ddft}) would lead to an artificial double-counting of
thermal fluctuations \cite{re:marconi99}. On the other hand, recent
studies \cite{re:archer04b}  have shown that deterministic 
DDFT dynamics governs the ensemble averaged density field 
$\rho({\bm r},t)$, while if the density field is temporally
coarse-grained --as is the assumption in PFC modeling-- 
the corresponding dynamic equation would then be
stochastic, but with a (unknown) coarse-grained free energy functional
instead of the equilibrium Helmholtz free energy functional used in
static DFT. In the current case of PFC modeling, quite drastic
approximations have been made to the DFT free energy functional
(particularly at the level of the direct correlation functions; see e.g.,
Eqs. (\ref{eq:C2}), (\ref{eq:C3}), and (\ref{eq:C_expan})), and hence
it could be argued that the incorporation of noise terms in the PFC
dynamics would be necessary and useful to capture the qualitative 
effects of fluctuations in phenomena such a homogeneous nucleation. 
In what follows, noise will be added to the PFC models studied above 
and the corresponding stochastic amplitude equations will be derived 
for both single component and binary systems. 
%From a practical point of view, it is 
%well known that stochastic fluctuations play a crucial role in many 
%dynamical phenomena, such as in homogeneous nucleation, and cannot 
%be ignored in these instances.

\subsection{Single-component PFC}
\label{sec:pure}

The stochastic DDFT equation for single-component systems is given by
Eq. (\ref{eq:ddft}) with a multiplicative noise term ${\bm \nabla} \cdot
[ \sqrt{\rho({\bm r},t)} {\bm \zeta}({\bm r},t) ]$ added to the
right-hand-side, 
where the noise field ${\bm \zeta}({\bm r},t)$ is determined by
(with $\Gamma_0 = 2 k_B T M$)
\begin{equation}
\langle {\bm \zeta}({\bm r},t) \rangle =0, \qquad 
\langle \zeta^{\mu}({\bm r},t) \zeta^{\nu}({\bm r'},t') \rangle =
\Gamma_0 \delta ({\bm r} - {\bm r'}) \delta (t-t') \delta^{\mu\nu}
\quad (\mu, \nu = x, y, z).
\label{eq:noise}
\end{equation}
The corresponding dynamic equation governing the rescaled
density field $n$ is similar to Eq. (\ref{eq:ddft_n}), i.e.,
$\partial n / \partial t = M' {\bm \nabla} \cdot [ (1+n) {\bm \nabla} 
{\delta {\cal F}} / {\delta n} ] + {\bm \nabla} \cdot
[ \sqrt{(1+n) / \rho_l} ~ {\bm \zeta} ]$. Adopting the lowest order
approximation as given in Sec. \ref{sec:ddft_pure}, we can write the
rescaled stochastic PFC equation as
\begin{equation}
\partial n / \partial t = \nabla^2 \left [ -\epsilon n + (\nabla^2 +
  q_0^2)^2 n + g_2 n^2 +n^3 \right ] + {\bm \nabla} \cdot {\bm \zeta},
\label{eq:pfc_noise}
\end{equation}
where the rescaled noise ${\bm \zeta}$ is also determined by
Eq. (\ref{eq:noise}) but with $\Gamma_0 = 2v/({B^x}^2 R^d \rho_l)$
(where $d$ is the dimensionality).

To derive the associated stochastic amplitude equations, we follow the
standard multiple scale approach in the limit of small $\epsilon$,
which leads to the expansion of density field $n$ in terms of the
zeroth-mode average density $n_0$ and complex amplitudes $A_j$ that
are varying on slow scales $(X,Y,T)$; see Eq. (\ref{eq:n_expan}).
Effects of external noise can be approximated via a projection
procedure used in hydrodynamic analysis
\cite{re:graham74,re:hohenberg92}. Based on the
fact that thermal noises originate from the fluctuations or random
motion of individual atoms/molecules at the microscopic
scales, we can project ${\bm \zeta}$ onto the base modes given in
Eq. (\ref{eq:n_expan}), i.e.,
\begin{equation}
{\bm \zeta} = {\bm \zeta}_0(X,Y,T)+\sum\limits_{j=1}^{3} 
{\bm \zeta}_{A_j}(X,Y,T) e^{i{\bm q}_j^0 \cdot {\bm r}} + {\rm c.c.},
\label{eq:zexpan}
\end{equation}
where 
\begin{eqnarray}
& \langle {\bm \zeta}_0 \rangle = \langle {\bm \zeta}_{A_j} \rangle =0,
\quad \langle {\bm \zeta}_{A_i} {\bm \zeta}_{A_j} \rangle = 
\langle {\bm \zeta}_0 {\bm \zeta}_{A_j} \rangle = \langle {\bm \zeta}_0 
{\bm \zeta}_{A_j}^* \rangle =0, & \nonumber\\
& \langle \zeta_0^{\mu} \zeta_0^{\nu} \rangle = \vartheta_0
\Gamma_0 \delta ({\bm r} - {\bm r'}) \delta (t-t') \delta^{\mu\nu}, \quad
\langle \zeta_{A_i}^{\mu} {\zeta_{A_j}^{\nu}}^* \rangle = \vartheta_i
\Gamma_0 \delta ({\bm r} - {\bm r'}) \delta (t-t') \delta_{ij}
\delta^{\mu\nu}, & \label{eq:zeta_nA}
\end{eqnarray}
(with $i,j=1,2,3; \mu, \nu =x,y$).
Here $\vartheta_i$ ($i=0,1,2,3$) is a constant determining the noise correlation
strength, which can be approximated as $\vartheta_i=\vartheta=1/7$ if
equal contribution from all modes in Eq. (\ref{eq:zexpan}) is assumed.
Thus the random noise term in Eq. (\ref{eq:pfc_noise}) is given by
\begin{equation}
{\bm \nabla} \cdot {\bm \zeta} = \sum\limits_{j=1}^{3} 
i {\bm q}_j^0 \cdot {\bm \zeta}_{A_j} e^{i{\bm q}_j^0 \cdot
  {\bm r}} + {\rm c.c.} + \epsilon^{1/2} \left [ \partial_X \zeta_0^x
  + \partial_Y \zeta_0^y + \sum\limits_{j=1}^{3} \left ( \partial_X
    \zeta_{A_j}^x + \partial_Y \zeta_{A_j}^y \right ) 
  e^{i{\bm q}_j^0 \cdot {\bm r}} + {\rm c.c.} \right ].
\label{eq:dzeta}
\end{equation}
In order to be relevant in the amplitude expansion, it is necessary that  
${\bm \nabla} \cdot {\bm \zeta} \sim {\cal O}(\epsilon^{3/2})$,
leading to ${\bm \zeta}_{A_j} \sim {\cal O}(\epsilon^{3/2})$ and hence
the noise intensity $\Gamma_0 \sim {\cal O}(\epsilon)$. The latter
yields ${\bm \zeta}_0 \sim {\cal O}(\epsilon^{3/2})$, which can be deduced from 
Eq. (\ref{eq:zeta_nA}).

Following the procedure of multiple scale expansion and
retaining the random force contribution to the lowest order, we can
derive the following stochastic amplitude equations
\begin{eqnarray}
&\partial A_j / \partial t =  - q_0^2 \delta {\cal F} / \delta A_j^* +
\zeta_j,& \label{eq:amplA}\\
&\partial n_0 / \partial t = \nabla^2 \delta {\cal F} / \delta n_0 +
{\bm \nabla} \cdot {\bm \zeta_0},& \label{eq:ampln0}
\end{eqnarray}
where ${\cal F}$ is the effective free energy of the 
single-component amplitude representation (see
Refs. \cite{re:huang08,re:yeon10,re:huang10} for the detailed form),
which is given by Eq. (\ref{eq:F}) with $\psi_0$ and
$\psi_j$ set to 0. Also, $\zeta_j = i {\bm q}_j^0 
\cdot {\bm \zeta}_{A_j}$ ($j=1,2,3$) and
\begin{eqnarray}
& \langle \zeta_j \rangle = \langle {\bm \zeta}_0 \rangle =0, \quad
\langle \zeta_i \zeta_j \rangle = \langle {\bm \zeta}_0 \zeta_j
\rangle = \langle {\bm \zeta}_0 \zeta_j^* \rangle =0, & \nonumber\\
& \langle \zeta_i \zeta_j^* \rangle = \vartheta_i q_0^2 \Gamma_0 
\delta ({\bm r} - {\bm r'}) \delta (t-t') \delta_{ij}, \quad
\langle \zeta_0^{\mu} \zeta_0^{\nu} \rangle = \vartheta_0
\Gamma_0 \delta ({\bm r} - {\bm r'}) \delta (t-t') \delta^{\mu\nu}. &
\label{eq:zeta}
\end{eqnarray}
The noise dynamics is then consistent with the dynamics of amplitude
representation, i.e., nonconserved dynamics for $A_j$ in
Eq. (\ref{eq:amplA}) and conserved one for $n_0$ in
Eq. (\ref{eq:ampln0}).

\subsection{Binary PFC}

Similar to the single-component system, based on
Eq. (\ref{eq:ddft_AB}) the stochastic DDFT equations for a binary
system can be written as
\begin{equation}
\frac{\partial \rho_A}{\partial t} = {\bm \nabla} \cdot \left [
M_A \rho_A {\bm \nabla} \frac{\delta {\cal F}}{\delta \rho_A}
+ \sqrt{\rho_A} {\bm \zeta_A} \right ], \qquad
\frac{\partial \rho_B}{\partial t} = {\bm \nabla} \cdot \left [
M_B \rho_B {\bm \nabla} \frac{\delta {\cal F}}{\delta \rho_B}
+ \sqrt{\rho_B} {\bm \zeta_B} \right ],
\label{eq:ddft_ABs}
\end{equation}
where for noises $(\alpha, \beta = A, B; \mu, \nu = x, y, z;
\Gamma_{\alpha} = 2 k_B T M_{\alpha})$,
\begin{equation}
\langle {\bm \zeta_i}({\bm r},t) \rangle =0, \qquad
\langle \zeta_{\alpha}^{\mu}({\bm r},t) \zeta_{\beta}^{\nu}
({\bm r'},t') \rangle = 
\Gamma_{\alpha} \delta ({\bm r} - {\bm r'}) \delta (t-t')
\delta_{\alpha \beta} \delta^{\mu \nu}.
\label{eq:noise_AB}
\end{equation}
From Eqs. (\ref{eq:ndN}) and (\ref{eq:pfc_ndN}) the dynamics equations
for $n$ and $\psi$ fields can be rewritten as
\begin{eqnarray}
&\partial n / \partial t =& M_1 {\cal D}_1 + M_2 {\cal D}_2+
{\bm \nabla} \cdot \left [ \sqrt{1+n} \left ( \sqrt{1+\psi} 
{\bm \zeta_A} + \sqrt{1-\psi} {\bm \zeta_B} \right ) \right ],
\nonumber\\
&\partial \psi / \partial t =& \frac{1}{1+n} \left \{
(M_2-M_1 \psi) {\cal D}_1 + (M_1-M_2 \psi) {\cal D}_2 
\right. \label{eq:npsi}\\
&& \left. + (1-\psi) {\bm \nabla} \cdot \left [ \sqrt{(1+n)(1+\psi)}
  {\bm \zeta_A} \right ] - (1+\psi) {\bm \nabla} \cdot \left [
  \sqrt{(1+n)(1-\psi)} {\bm \zeta_B} \right ] \right \},
\nonumber
\end{eqnarray}
where we have rescaled ${\bm \zeta_{A(B)}} \rightarrow {\bm
  \zeta_{A(B)}}/\sqrt{2\rho_l}$. Following the procedure given in
Sec. \ref{sec:ddft_binary} and only retaining the lowest order noise
terms, we can derive the rescaled stochastic binary PFC equations as
\begin{equation}
\partial n / \partial t = {\cal D}_1 + m {\cal D}_2
 + {\bm \nabla} \cdot {\bm \zeta}_n, \quad
\partial \psi / \partial t = m {\cal D}_1 + {\cal D}_2
+ {\bm \nabla} \cdot {\bm \zeta}_{\psi},
\label{eq:bpfc}
\end{equation}
where the expressions of ${\cal D}_1$ and ${\cal D}_2$ have been given
in Eqs. (\ref{eq:D12})--(\ref{eq:F_eff}). The noise terms are defined
by
\begin{equation}
{\bm \zeta}_n = {\bm \zeta}_A + {\bm \zeta}_B, \qquad
{\bm \zeta}_{\psi} = {\bm \zeta}_A - {\bm \zeta}_B,
\end{equation}
where $\zeta_{A(B)}$ also obeys Eq. (\ref{eq:noise_AB}), although with 
$\Gamma_{\alpha}=k_BTM_{\alpha}v/(M_1{B_0^x}^2R^d\rho_l)$ due to the
rescaling, and
\begin{eqnarray}
& \langle {\bm \zeta}_n \rangle = \langle {\bm \zeta}_{\psi} \rangle =0,
\quad \langle \zeta_n^{\mu} \zeta_{\psi}^{\nu} \rangle =
(\Gamma_A-\Gamma_B) \delta ({\bm r} - {\bm r'}) \delta (t-t')
\delta^{\mu\nu},& \nonumber\\
& \langle \zeta_n^{\mu} \zeta_n^{\nu} \rangle = 
\langle \zeta_{\psi}^{\mu} \zeta_{\psi}^{\nu} \rangle =
(\Gamma_A+\Gamma_B) \delta ({\bm r} - {\bm r'}) \delta (t-t')
\delta^{\mu\nu},& \label{eq:zeta_npsi}
\end{eqnarray}
with $\Gamma_A+\Gamma_B = 2v/({B_0^x}^2R^d\rho_l)$ and
$\Gamma_A-\Gamma_B = m (\Gamma_A+\Gamma_B) =
2mv/({B_0^x}^2R^d\rho_l)$. 

Using the multiple scale approach, we can expand the density field $n$
according to Eq. (\ref{eq:n_expan}) while assuming the concentration
field as slowly varying, $\psi = \psi_0(X,Y,T)$ (that is, keeping only
the zeroth mode and neglecting the higher-order contributions from
$\psi_j$ in Eq. (\ref{eq:psi_expan}), as discussed in
Sec. \ref{sec:hybrid}). Similar to the single-component case,
the projection of noises can be given by
\begin{equation}
{\bm \zeta}_n = {\bm \zeta}_0(X,Y,T)+\sum\limits_{j=1}^{3} 
{\bm \zeta}_{A_j}(X,Y,T) e^{i{\bm q}_j^0 \cdot {\bm r}} + {\rm c.c.},
\quad {\bm \zeta}_{\psi} = {\bm \zeta}_{\psi}(X,Y,T).
\label{eq:bzexpan}
\end{equation}
Thus the expression of ${\bm \nabla} \cdot {\bm \zeta}_n$ is the same
as Eq. (\ref{eq:dzeta}), while ${\bm \nabla} \cdot {\bm \zeta}_{\psi}
= \epsilon^{1/2} (\partial_X \zeta_{\psi}^x + \partial_Y \zeta_{\psi}^y)$.
Also we can estimate ${\bm \zeta}_{A_j}, {\bm \zeta}_0, {\bm \zeta}_{\psi}
\sim {\cal O}(\epsilon^{3/2})$ and $\Gamma_A, \Gamma_B \sim {\cal
  O}(\epsilon)$. 

The stochastic amplitude equations for binary PFC model can then be
derived, i.e.,
\begin{eqnarray}
\partial A_j / \partial t &=& - q_0^2 (1-m^2) \frac{\delta {\cal F}}
{\delta \A_j^*} + \zeta_j, \label{eq:A_noise} \\
\partial n_0 / \partial t &=& \nabla^2 \frac{\delta {\cal F}}
{\delta n_0} + m \nabla^2 \frac{\delta {\cal F}}{\delta \psi_0}
+ {\bm \nabla} \cdot {\bm \zeta}_0, \label{eq:n0_noise} \\
\partial \psi_0 / \partial t &=& m \nabla^2 \frac{\delta {\cal F}}
{\delta n_0} + \nabla^2 \frac{\delta {\cal F}}{\delta \psi_0}
+ {\bm \nabla} \cdot {\bm \zeta}_{\psi_0}, \label{eq:psi0_noise}
\end{eqnarray}
where the deterministic parts have been obtained in
Sec. \ref{sec:hybrid}; see Eqs. (\ref{eq:n0}), (\ref{eq:psi0}), and
(\ref{eq:Aj0}), as well as Eq. (\ref{eq:F}) for the effective
potential ${\cal F}$. For the noise terms, 
$\zeta_j = i {\bm q}_j^0 \cdot {\bm \zeta}_{A_j}$ ($j=1,2,3$), and
\begin{eqnarray}
& \langle \zeta_j \rangle = \langle {\bm \zeta}_0 \rangle = 
\langle {\bm \zeta}_{\psi_0} \rangle = 0, \quad
\langle \zeta_i \zeta_j \rangle = \langle {\bm \zeta}_0 \zeta_j
\rangle = \langle {\bm \zeta}_0 \zeta_j^* \rangle =
\langle {\bm \zeta}_{\psi_0} \zeta_j \rangle = \langle {\bm
  \zeta}_{\psi_0} \zeta_j^* \rangle = 0, & \nonumber\\
& \langle \zeta_i \zeta_j^* \rangle = \vartheta_i q_0^2 (\Gamma_A +
\Gamma_B) \delta ({\bm r} - {\bm r'}) \delta (t-t') \delta_{ij}, \quad
\langle \zeta_0^{\mu} \zeta_0^{\nu} \rangle = \vartheta_0
(\Gamma_A + \Gamma_B) \delta ({\bm r} - {\bm r'}) \delta (t-t')
\delta^{\mu\nu}, & \label{eq:bzeta}\\
& \langle \zeta_{\psi_0}^{\mu} \zeta_{\psi_0}^{\nu} \rangle =
(\Gamma_A + \Gamma_B) \delta ({\bm r} - {\bm r'}) \delta (t-t')
\delta^{\mu\nu}, \quad 
\langle \zeta_{\psi_0}^{\mu} \zeta_0^{\nu} \rangle =
(\Gamma_A - \Gamma_B) \delta ({\bm r} - {\bm r'}) \delta (t-t')
\delta^{\mu\nu},& \nonumber
\end{eqnarray}
with $i,j=1,2,3$ and $\mu, \nu = x,y$. If assuming $\Gamma_A \simeq
\Gamma_B$ (for equal mobility $M_A \simeq M_B$ and $m \simeq 0$),
i.e., with almost the same noise/fluctuation intensity for A and B
components, we have $\langle \zeta_{\psi_0}^{\mu} \zeta_0^{\nu} \rangle
\simeq 0$ and hence all noise terms ($\zeta_j$, ${\bm \zeta}_0$, ${\bm
  \zeta}_{\psi_0}$) can be treated independently. However, for the case
of different mobilities ($M_A \neq M_B$ and $m \neq 0$), we get
$\langle \zeta_{\psi_0}^{\mu} \zeta_0^{\nu} \rangle \neq 0$, and hence
noises ${\bm \zeta}_0$ and ${\bm \zeta}_{\psi_0}$ for 0th-mode density fields
$n_0$ and $\psi_0$ are then correlated. Similar results can be obtained
for noises ${\bm \zeta}_n$ and ${\bm \zeta}_{\psi}$ in the stochastic
PFC equations (\ref{eq:npsi}) and (\ref{eq:zeta_npsi}).

\section{Applications in Alloy Heterostructures}
\label{sec:appl}

As discussed in the introduction, the PFC model and the 
corresponding amplitude equations have applied to the study 
of a wide variety of phenomena involved in material processing and
microstructure evolution.  In this section we will illustrate how 
the amplitude equations derived in the preceding sections can 
be employed to examine the effect of surface segregation and alloy
intermixing. Alloy intermixing is known to play an important role in
the growth and processing of material heterostructures, including
morphological and compositional profiles and the associated sample
optoelectronic properties and functionality. Recent intensive studies
on thin film epitaxy and atomic deposition have shown the
important effects of intermixing on nanostructure
self-assembly. Typical examples include InAs(InGaAs)/GaAs(001)
\cite{re:moison89,re:walther01,re:cederberg07} or Ge(SiGe)/Si(001) 
\cite{re:walther97,re:denker05}
heteroepitaxy that has been investigated extensively (particularly the
intermixing-caused alloying of wetting layers and quantum dots), and
the interlayer diffusion in semiconductor multilayers or superlattices
such as InP/InGaAs \cite{re:gerling01}, GaAs/GaSb \cite{re:dorin03},
and GaAs/InAs \cite{re:pearson04}.
An important phenomenon in these epitaxial layers is the occurrence of
surface segregation, in which an enrichment of one of the film
components at a surface or interface region occurs.  This has 
been observed in a variety of material systems including III-V and
II-VI semiconductor heterostructures \cite{re:moison89,re:walther01,%
re:cederberg07,re:walther97,re:denker05,re:gerling01,re:dorin03,re:pearson04}.
To address these complicated phenomena and effects, the
basic processes and mechanisms of intra- and inter-layer diffusion at
nearly-planar interfaces as well as their coupling with material
processing and growth parameters needs to be clarified. 
%which serve as a basis for further modeling and understanding
%the formation of more complicated surface or interface structures.

In light of the above observations, the focus of this section is on 
heterostructures of a nearly-planar interface, for both
lattice-matched and strained epitaxial layers. For
layers stressed due to lattice mismatch, the configurations studied
here are metastable in nature, and our results will be used for further
studies of the associated later-stage nanostructure evolution (e.g.,
quantum dots), which  will be presented elsewhere. For such film geometry
with a planar interface, it can be assumed that both morphological and
compositional profiles along the lateral direction are approximately
uniform or homogeneous (at least metastably), and hence these
structural profiles vary only along the direction ($y$) normal to the
interface. An advantage of the amplitude equation representation
of the PFC model developed above is that the system of interest can then be mapped onto
an effective one-dimensional (1D) description, as will be shown below.

\subsection{Effective 1D model system with elasticity}

To address the elasticity incorporated in the amplitude
equation formalism, it is useful to note that the structural amplitudes
can be written as
\begin{equation}
A_j = A_j' e^{i {\bm q}_j^0 \cdot {\bm u}} \quad (j=1,2,3),
\end{equation} 
where for 2D hexagonal structure ${\bm q}_j^0$ are the three basic
wave vectors given in Sec. \ref{sec:ampl} and ${\bm u} = \delta_0 (x
\hat{x} + y \hat{y})$ describes the bulk compression or dilation. The
effective free energy ${\cal F}$ in Eq. (\ref{eq:F}) can be rewritten as
(neglecting the higher order contributions from $\psi_j$ and
approximating ${\cal L}_j \simeq -q_0^2$)
\begin{eqnarray}
& {\cal F} = \int d{\bm r} & \left \{ - \frac{1}{2} \epsilon n_0^2 +
  \frac{1}{2} \left [ \left ( \nabla^2 + q_0^2 \right ) n_0 \right ]^2
  + \frac{1}{3} g_2 n_0^3 + \frac{1}{4} n_0^4
  + \left ( - \epsilon + 3 n_0^2 + 2 g_2 n_0 + g \psi_0^2 \right )
  \sum_j |\A_j'|^2 \right. \nonumber\\
&& + \sum_j \left | {\cal G}_j' \A_j' \right |^2 + \frac{3}{2}
\sum_j |\A_j'|^4 + (6n_0 + 2g_2) \left ( \prod_j \A_j' +
  {\rm c.c.} \right ) + 6 \sum_{j<k} |\A_j'|^2 |\A_k'|^2 \nonumber\\
&& + \frac{1}{2} w_0 \psi_0^2 + \frac{1}{2} K_0 \left | \nabla
  \psi_0 \right |^2 + \frac{1}{4} u_0 \psi_0^4 
+ \frac{1}{2} g n_0^2 \psi_0^2 + v_1 n_0 \psi_0^2  \nonumber\\
&& \left. + \etaa \left [ \psi_0 n_0 \left ( \nabla^2 + \nabla^4 \right )
  n_0 - q_0^2 \psi_0 \left ( \sum_j {\A_j'}^* {\cal G}_j' \A_j'
    + {\rm c.c.} \right ) \right ] \right \}, \label{eq:F'}
\end{eqnarray}
where
\begin{equation}
{\cal G}_j' = \nabla^2 + 2i \left ({\bm \delta}_j + {\bm q}_j^0 \right
) \cdot {\bm \nabla} - |{\bm \delta}_j|^2 - 2 {\bm q}_j^0 \cdot {\bm
  \delta}_j, 
\end{equation}
with $\delta_1 = -\delta_x \hat{x} - \delta_y \hat{y}/2$,
$\delta_2 = \delta_y \hat{y}$,
$\delta_3 = \delta_x \hat{x} - \delta_y \hat{y}/2$,
$\delta_x = \sqrt{3} q_0 \delta_0 /2$, and $\delta_y = q_0 \delta_0$.
The corresponding dynamic equations for $A_j'$, $n_0$, and $\psi_0$
are still governed by Eqs. (\ref{eq:A_noise})--(\ref{eq:psi0_noise}),
although with $A_j$ replaced by $A_j'$.
In mechanical equilibrium, we can assume that $A_j' \simeq
A$, i.e., $A_j \simeq A \exp(i {\bm q}_j^0 \cdot {\bm u})$ where $A$ is 
a constant. Minimizing the effective free energy ${\cal F}$ with
respect to A yields the equilibrium value
$\delta_0^{\rm eq} = -1 + \sqrt{1-2\alpha_0\psi_0} \simeq -
\alpha_0 \psi_0$ to lowest order. This leads to the equilibrium wave
number $q_{\rm eq} = (1+\delta_0^{\rm eq}) q_0 =
\sqrt{1-2\alpha_0\psi_0} ~q_0$ (where $\alpha_0$ is the
rescaled solute expansion coefficient defined in
Sec. \ref{sec:ddft_binary}), and the equilibrium amplitude
\begin{equation}
A = \frac{1}{15} \left \{ -(3n_0+g_2) + \sqrt{(3n_0+g_2)^2 - 15 \left
      [ -\epsilon + q_0^4 (\delta_0^2 + 2\delta_0)(\delta_0^2 +
      2\delta_0+ 4\alpha_0 \psi_0) + n_0 (3n_0+2g_2) + g\psi_0^2
    \right ]} \right \}. \label{eq:A_eq}
\end{equation}
The elastic constants (rescaled) are then given by
$C_{11}=C_{22}=9A^2$, $C_{12}=C_{44}=C_{11}/3=3A^2$, and Young's
modulus $E=8A^2$ \cite{re:elder04,re:elder07,re:elder10}. 

For the dynamics of a heterostructure configuration with nearly-planar interface
(either liquid-solid or solid-solid), we can assume that $A_j'(x,y,t) \simeq A_j^0(y,t)$, 
$n_0(x,y,t) \simeq n_0^0(y,t)$, and $\psi_0(x,y,t) \simeq \psi_0^0(y,t)$, resulting in an
effective 1D description of the system. The dynamics of the
amplitude equations then become
\begin{eqnarray}
\partial n_0^0 / \partial t &=& \partial_y^2 \frac{\delta {\cal F}}
{\delta n_0^0} + m \partial_y^2 \frac{\delta {\cal F}}{\delta \psi_0^0},
\label{eq:n0_1D} \\
\partial \psi_0^0 / \partial t &=& m \partial_y^2 \frac{\delta {\cal F}}
{\delta n_0^0} + \partial_y^2 \frac{\delta {\cal F}}{\delta \psi_0^0},
\label{eq:psi0_1D} \\
\partial A_j^0 / \partial t &=& - q_0^2 (1-m^2) 
\frac{\delta {\cal F}}{\delta {\A_j^0}^*},
\label{eq:A_1D}
\end{eqnarray}
where
\begin{eqnarray}
\frac{\delta {\cal F}}{\delta n_0^0} &=& \left [ -\epsilon + \left
    ( \partial_y^2 + q_0^2 \right )^2 \right ] n_0^0 + g_2 {n_0^0}^2 +
{n_0^0}^3 + (6n_0^0+2g_2) \sum_j |A_j^0|^2 + 6 \left (\prod_j A_j^0 +
  {\rm c.c.} \right ) + (g n_0^0 + v_1) {\psi_0^0}^2 \nonumber\\
&&+ 2\alpha_0 \left [ \psi_0^0 \left ( \partial_y^2 + \partial_y^4
  \right ) n_0^0 + \left ( \partial_y^2 + \partial_y^4 \right ) \left
    ( n_0^0 \psi_0^0 \right ) \right ], \label{eq:dF_dn0} \\
\frac{\delta {\cal F}}{\delta \psi_0^0} &=& (w_0-K_0\partial_y^2)
\psi_0^0 + u_0 {\psi_0^0}^3 + g \left ( {n_0^0}^2 + 2
  \sum_j |A_j^0|^2 \right ) \psi_0^0 + 2v_1 n_0^0 \psi_0^0 \nonumber\\
&&+ 2 \alpha_0\left [ n_0^0 \left ( \partial_y^2 + \partial_y^4 \right )
  n_0^0 - q_0^2 \sum_j \left ( {A_j^0}^* {\cal G}_j^0 A_j^0 + {\rm
      c.c.} \right ) \right ], \label{eq:dF_dpsi0} \\
\frac{\delta {\cal F}}{\delta {\A_j^0}^*} &=& \left [ -\epsilon +
  {{\cal G}_j^0}^2 + 2g_2 n_0^0 + 3{n_0^0}^2 + g{\psi_0^0}^2 \right ]
\A_j^0 + 3 \A_j^0 \left [ |\A_j^0|^2 + 2 \sum_{k,l \neq j}^{k<l} \left
    ( |A_k^0|^2 + |A_l^0|^2 \right ) \right ] \nonumber\\
&& + (6n_0^0 + 2g_2) \prod_{k \neq j} {A_k^0}^* - 2\alpha_0 q_0^2
\left [ \psi_0^0 {\cal G}_j^0 A_j^0 + {\cal G}_j^0 \left ( \psi_0^0
    A_j^0 \right ) \right ], \label{eq:dF_dAj}
%\left. \frac{\delta {\cal F}}{\delta \psi_j^*} \right |_{\psi_j=0} &=&
%2g\psi_0^0 \left ( n_0^0 A_j^0 + \prod_{k \neq j} {A_k^0}^* \right )
%+ 2 v_1 \psi_0^0 A_j^0 + 2\alpha_0 \left [ A_j^0 \left ( \partial_y^2
%    + \partial_y^4 \right ) n_0^0 -q_0^2 n_0^0 {\cal G}_j^0 A_j^0
%- q_0^2 \sum_{k \neq l \neq j} {A_k^0}^* {{\cal G}_l^0}^* {A_l^0}^*
%\right ], \label{eq:dF_dpsij}
\end{eqnarray}
with
\begin{equation}
{\cal G}_j^0 = \partial_y^2 + 2i \left (\delta_{jy} + q_{jy}^0 \right )
\partial_y - |{\bm \delta}_j|^2 - 2 {\bm q}_j^0 \cdot {\bm \delta}_j.
\end{equation}

For coherent strained alloy layers, which are of great interest in
materials growth, the solid layer is strained with respect to a
substrate and subjected to an epitaxial condition 
$q_x=q_x^{\rm sub}=(\sqrt{3}/2) q_0 (1+\delta_0^{\rm sub})$ (with
``sub'' referring to the substrate). 
The wavenumber $q_y$ along the vertical or layer
growth direction $y$ is determined by the lattice elastic relaxation
(or Poisson relaxation in continuum elasticity theory). The system is
thus governed by the above amplitude equations
(\ref{eq:n0_1D})--(\ref{eq:dF_dAj}), but with $\delta_0$ fixed by the
corresponding elasticity quantity $\delta_0^{\rm sub}$ of the
substrate (and thus $\delta_x=\sqrt{3} q_0 \delta_0^{\rm sub}
/2$ and $\delta_y=q_0 \delta_0^{\rm sub}$). The vertical strain
relaxation (Poisson relaxation) can be 
determined from the phase of complex amplitudes $A_j^0$.
Furthermore, the misfit strain $\varepsilon_m$ of such a solid layer is
given by
\begin{equation}
\varepsilon_m = \frac{R_{\rm eq} - R}{R} =
\frac{q_x}{q_{x,\rm eq}}-1 = \frac{\delta_0 - \delta_0^{\rm eq}}
{1+\delta_0^{\rm eq}},
\label{eq:misfit}
\end{equation}
where $R$ and $q_x$ are lateral lattice spacing and wavenumber of the
strained layer, and $R_{\rm eq}$, $q_{x,\rm eq}$, and $\delta_0^{\rm eq}$ 
are for the corresponding stress-free, equilibrium bulk state.

For the systems studied here the model parameters are chosen such that
no phase separation or spinodal decomposition can occur in the bulk
of each solid or liquid region. The corresponding conditions on the parameters 
that assure this are derived via a linear stability analysis of the amplitude equations.
Following standard procedures, we substitute the expansion
$n_0=\bar{n}_0+\hat{n}_0$, $\psi_0=\bar{\psi}_0+\hat{\psi}_0$,
and $A_j=\bar{A}_j+\hat{A}_j$ into Eqs. (\ref{eq:n0}), (\ref{eq:psi0})
and (\ref{eq:Aj0}), obtain the linearized 
evolution equations for the perturbed quantities $\hat{n}_0$,
$\hat{\psi}_0$, and $\hat{A}_j$, and calculate the associated
perturbation growth rates. The corresponding results are 
complicated due to the coupling between the evolution equations of all
three perturbed quantities. To estimate the conditions for phase
separation, here we simply assume that $\hat{n}_0,
\hat{A}_j \sim 0$, and only study the stability of concentration
field. To first order of $\hat{\psi}_0$ we have
\begin{equation}
\partial \hat{\psi}_0 / \partial t \simeq \nabla^2 \left \{ -K_0
  \nabla^2 + w_0 + 3u_0 \bar{\psi}_0^2 + g \bar{n}_0^2 + 2v_1
  \bar{n}_0 + 2g \sum_j |\bar{A}_j|^2 + m \left [ \etaa \bar{n}_0
    (\nabla^2+\nabla^4) + 2g\bar{n}_0\bar{\psi}_0 + 2v_1 \bar{\psi}_0
  \right ] \right \} \hat{\psi}_0. \label{eq:psi0_lin}
\end{equation}
In Fourier space, the perturbation growth rate $\sigma(q)$ is then given by
\begin{equation}
\sigma = -q^2 \left [ 2 m \alpha_0 \bar{n}_0 q^4 + (K_0 - 2m\alpha_0
  \bar{n}_0) q^2 + w_{\rm eff} \right ],
\end{equation}
where
\begin{equation}
w_{\rm eff} = w_0 + 3u_0 \bar{\psi}_0^2 + g \bar{n}_0^2 + 2v_1
\bar{n}_0 + 2g \sum_j |\bar{A}_j|^2 + 2m (g\bar{n}_0+v_1) \bar{\psi}_0.
\label{eq:w_eff}
\end{equation}
If $w_{\rm eff}<0$, an instability of the homogeneous alloy
occurs, leading to spinodal decomposition or phase separation of alloy
components. The characteristic wave number (for maximum perturbation
growth rate) is then given by $q_{\rm max}^2 = [ \sqrt{(K_0 -
  2m\alpha_0\bar{n}_0)^2 - 6m\alpha_0\bar{n}_0 w_{\rm eff}} - (K_0 -
2m\alpha_0\bar{n}_0) ] / (6m\alpha_0\bar{n}_0)$ if $m, \alpha_0,
\bar{n}_0 \neq 0$, or $q_{\rm max}^2 = - w_{\rm eff}/(2K_0)$ if one of
$m, \alpha_0, \bar{n}_0 =0$.

For the heterostructural systems presented below and the parameters
chosen, the condition $w_{\rm eff}>0$ is always satisfied in the bulk phases,
keeping homogeneous concentration profile within each layer.
Concentration heterogeneity may occur across the system configuration,
which however is due to the effect of interfaces or due to composition
overshooting, a phenomenon caused by alloy intermixing that will be
discussed in detailed below.

\subsection{Results: Equilibrium profiles and layer growth}

Equations (\ref{eq:n0_1D})--(\ref{eq:dF_dAj}) were solved numerically 
using a pseudospectral method and an exponential propagation scheme
for time integration of stiff equations \cite{re:friesner89,re:cross94b}.
Results of the corresponding morphological and compositional 1D
profiles are shown in Figs. \ref{fig:ssl_P}--\ref{fig:sl_f_m}, for two types 
of configurations of liquid-solid-solid and liquid-solid coexistence or 
growth. For the simulations shown here we choose a time step $\Delta t =1$, 
which can be made as large as this due to the numerical scheme we used; 
The numerical grid spacing used is $\Delta y = \lambda_0/8$ (where 
$\lambda_0 = 2\pi /q_0$). To emulate  a liquid-solid (or liquid-solid-solid) 
heterostructure and apply periodic boundary conditions in the
numerical calculation, the initial configuration is set as two (or four)
symmetric interfaces located at $y=L_y/4$ and $3L_y/4$ (or $y=L_y/6$,
$L_y/3$, $2L_y/3$ and $5L_y/6$), separating different liquid or solid
regions. These interfaces need to be set sufficiently far
apart from each other to avoid any interface coupling and the
artifacts of finite size effects. For results shown below we choose
the 1D system size perpendicular to the interfaces as $L_y=2048 \Delta
y$, with similar results obtained in calculations up to $L_y=8192
\Delta y$. Also, the parameters used in the amplitude equations 
are based on the phase diagrams given in Ref. \cite{re:elder10} showing
liquid-solid and solid-solid coexistence, i.e., $(g,g_2,u_0,K_0,v_1)
=(-1.8,-0.6,4,1,0)$, $w_0=0.008$ or $0.088$, $\alpha_0=0.3$ or $0$,
and $\epsilon=\pm 0.02$.

\subsubsection{Liquid-solid and liquid-solid-solid coexistence}

\begin{figure}
\centerline{
\includegraphics[height=3.3in]{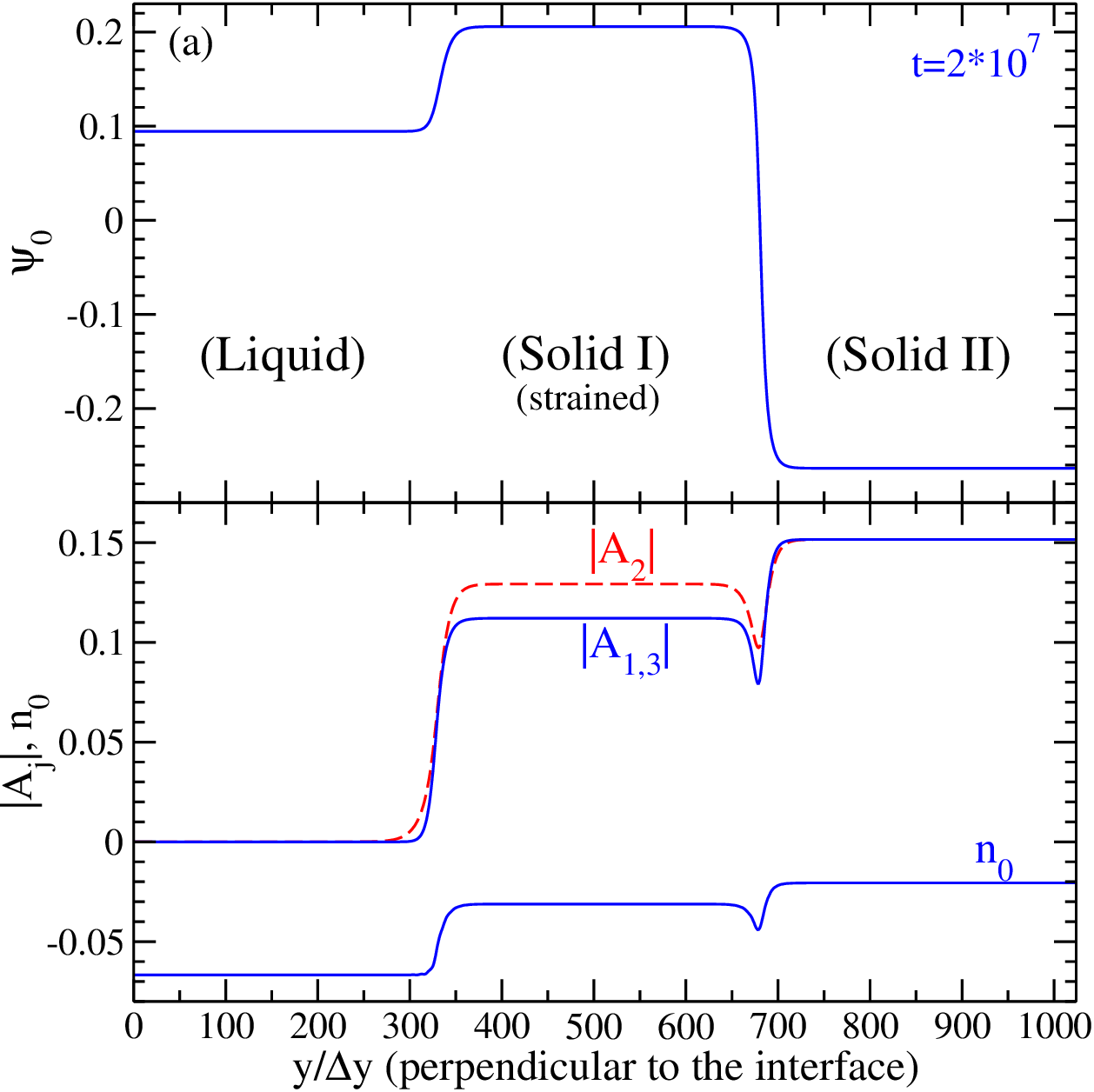} \hskip 5pt
\includegraphics[height=3.3in]{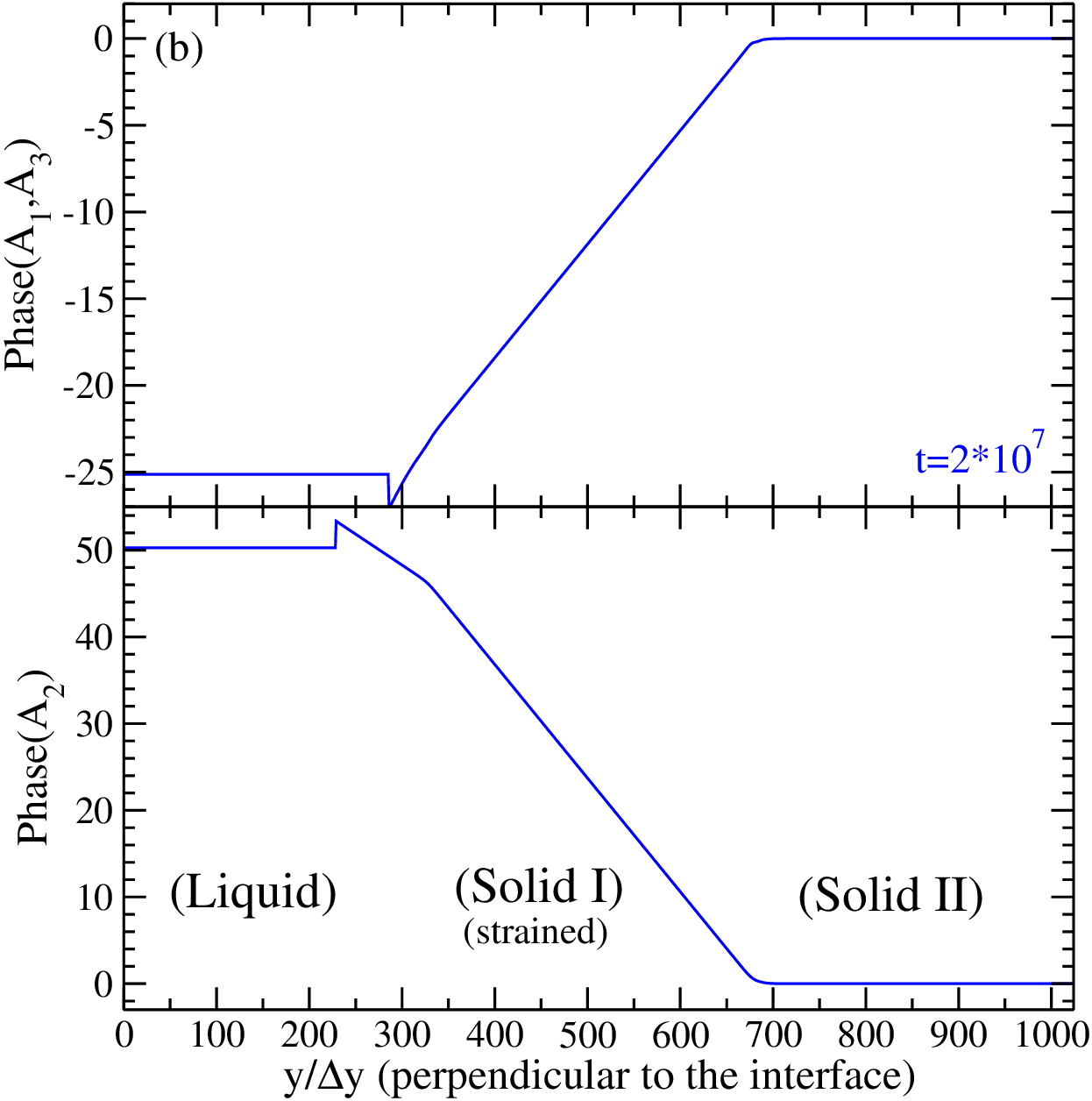}}
\caption{Liquid-solid-solid coexistence profile calculated from the
  amplitude equations, as characterized by (a) the composition field
  $\psi_0$, amplitudes $|A_j|$, and the average density field $n_0$,
  and (b) the phases of amplitudes $A_j$. The parameters are set as
  $\epsilon=0.02$, $\alpha_0=0.3$, $(g,g_2,u_0,K_0,w_0)
  =(-1.8,-0.6,4,1,0.008)$, and the time at $t=2\times 10^7$. Solid I is
  strained with respect to the substrate solid II.}
\label{fig:ssl_P}
\end{figure}

The equilibrium profile for a liquid-solid(I)-solid(II) coexistence is
given in Fig. \ref{fig:ssl_P} (with time corresponding to $t=2 \times 10^7$).
To obtain the liquid-solid-solid coexistence, we use $\epsilon=0.02$,
$\alpha_0=0.3$, and $w_0=0.008$ (from the eutectic phase diagram in
Ref. \cite{re:elder10}), set the initial length ratio of
liquid:solid(I):solid(II) as 1/3:1/3:1/3, and let all of $\psi_0^0$,
$A_j^0$ and $n_0^0$ evolve with time until a stationary state is
reached. Solid II is treated as a substrate (unstrained), and hence in
the amplitude equations (\ref{eq:n0_1D})--(\ref{eq:dF_dAj}) we set
$\delta_0=\delta_0^{\rm II}=-1 + \sqrt{1-2\alpha_0\psi_0^{\rm II}}$.
Due to nonzero solute expansion coefficient $\alpha_0$, i.e.,
different atomic sizes of A and B alloy components, solid I is
strained (with misfit $\varepsilon_m$ with respect to the substrate
(solid II) being $14.9\%$ for the parameters of
Fig. \ref{fig:ssl_P}). This is consistent with the numerical results
in Fig. \ref{fig:ssl_P}b, showing zero phase of amplitudes $A_j$
within unstrained solid II and a linear dependence of phase on
position $y$ in the bulk of solid I.
For comparison, the magnitude of lattice misfit between III-V or II-VI
layers is around 0 to $5\%$ (e.g., $\varepsilon_m=4.2\%$ for Ge/Si and
less for Si$_x$Ge$_{1-x}$/Si$_y$Ge$_{1-y}$), while the lattice
mismatch for III-V Nitride heteroepitaxial films or III-V/Si
heterostructures could reach 10\% or more (e.g.,
$\varepsilon_m=11.5\%$ for InAs/Si).

\begin{figure}
\centerline{
\includegraphics[height=3.3in]{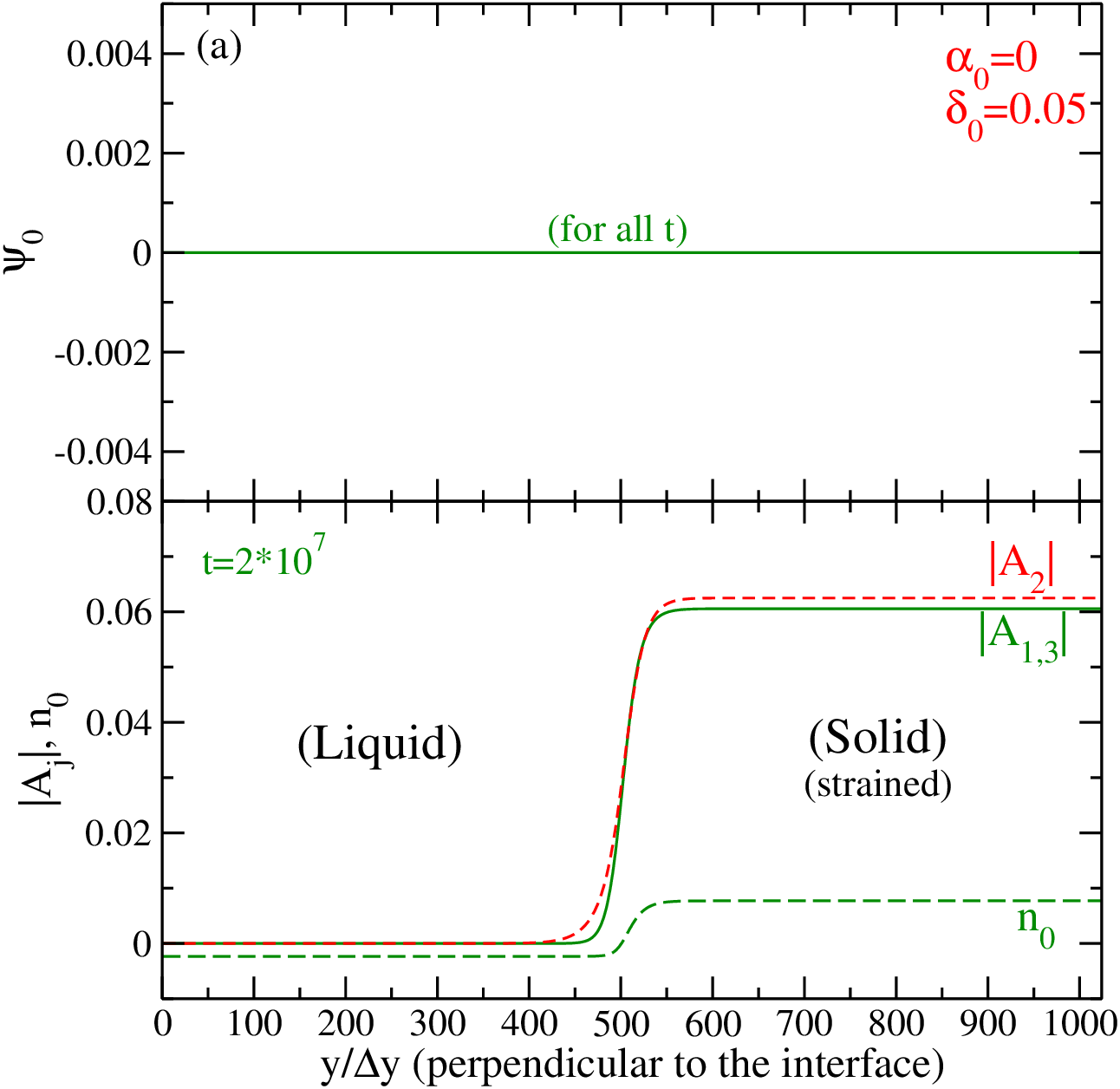} \hskip 5pt
\includegraphics[height=3.3in]{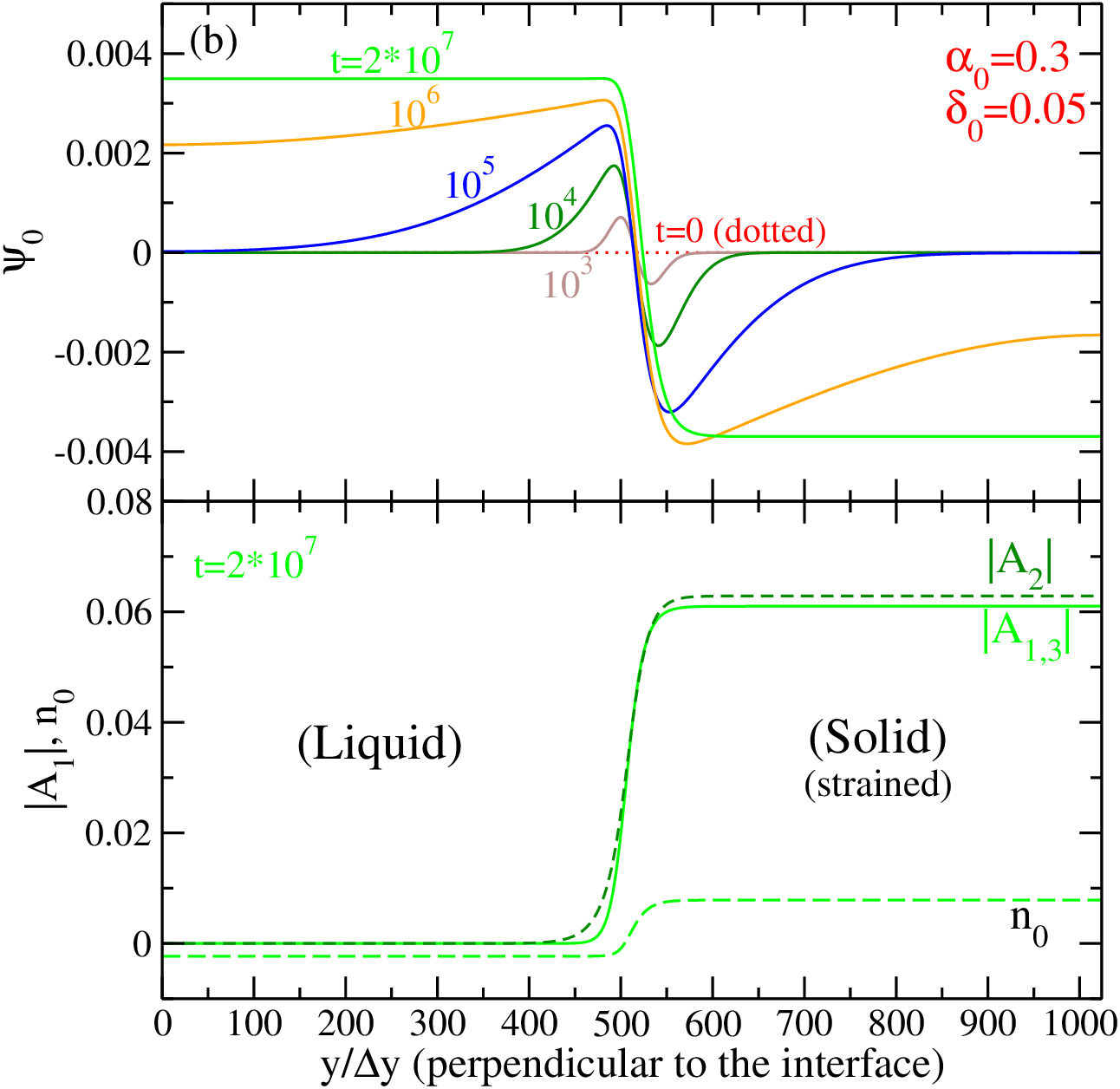}}
\caption{Liquid-solid coexistence profiles, for $\alpha_0=0$ (a) and
  $0.3$ (b). The other parameters are the same as
  Fig. \ref{fig:ssl_P}, except for $\epsilon=-0.02$, $w_0=0.088$, and
  initially $\psi_0=0$. The solid region has a misfit strain of around
  $5\%$ due to $\delta_0=0.05$ set in the amplitude equations.}
\label{fig:slPeq}
\end{figure}

For a liquid-solid heterostructural configuration, to determine the
coexistence state we choose similar parameters except for $w_0=0.088$,
$\epsilon=-0.02$, and initially $\psi_0=0$ in the whole system. This
corresponds to the single solid phase region (no solid-solid
coexistence, only liquid-solid) in the phase diagram \cite{re:elder10}.
To make the solid strained, we set $\delta_0=0.05$ as given by an
external condition (i.e., a substrate), and thus from
Eq. (\ref{eq:misfit}) the misfit strain in the solid here is about
$5\%$. The results for $\alpha_0=0$ and $0.3$ are given
in Figs. \ref{fig:slPeq}a and \ref{fig:slPeq}b respectively,
including the equilibrium profiles (up to $t=2\times 10^7$) and the
process of time evolution. As expected, for $\alpha_0=0$ (equal atomic
size of alloy components) the concentration field $\psi_0$
remains at 0 all the time, as seen in Fig. \ref{fig:slPeq}a. However,
for $\alpha_0=0.3$ the initial $\psi_0=0$ profile splits at the
liquid-solid interface (see Fig. \ref{fig:slPeq}b).  
For the parameters used here, $\alpha_0>0$ (with size of atom A larger
than that of atom B) and misfit 
$\varepsilon_m>0$ (compressed solid), and thus the solid would prefer
to have more smaller atoms B (with $\psi_0<0$), leading to a ``dip''
on the solid side of the compositional interface; due to the conservation
law on the field $\psi_0$, a ``bump'' of $\psi_0>0$ (more larger atoms A)
appears on the other side via layer interdiffusion or
alloy intermixing. As a result of atomic
diffusion, such ``dip'' and ``bump'' will spread out into the bulk phases as
time increases, leading to a positive/negative $\psi_0$ equilibrium
profile of liquid-solid coexistence, as shown in Fig. \ref{fig:slPeq}b.

\begin{figure}
\centerline{
\includegraphics[height=3.3in]{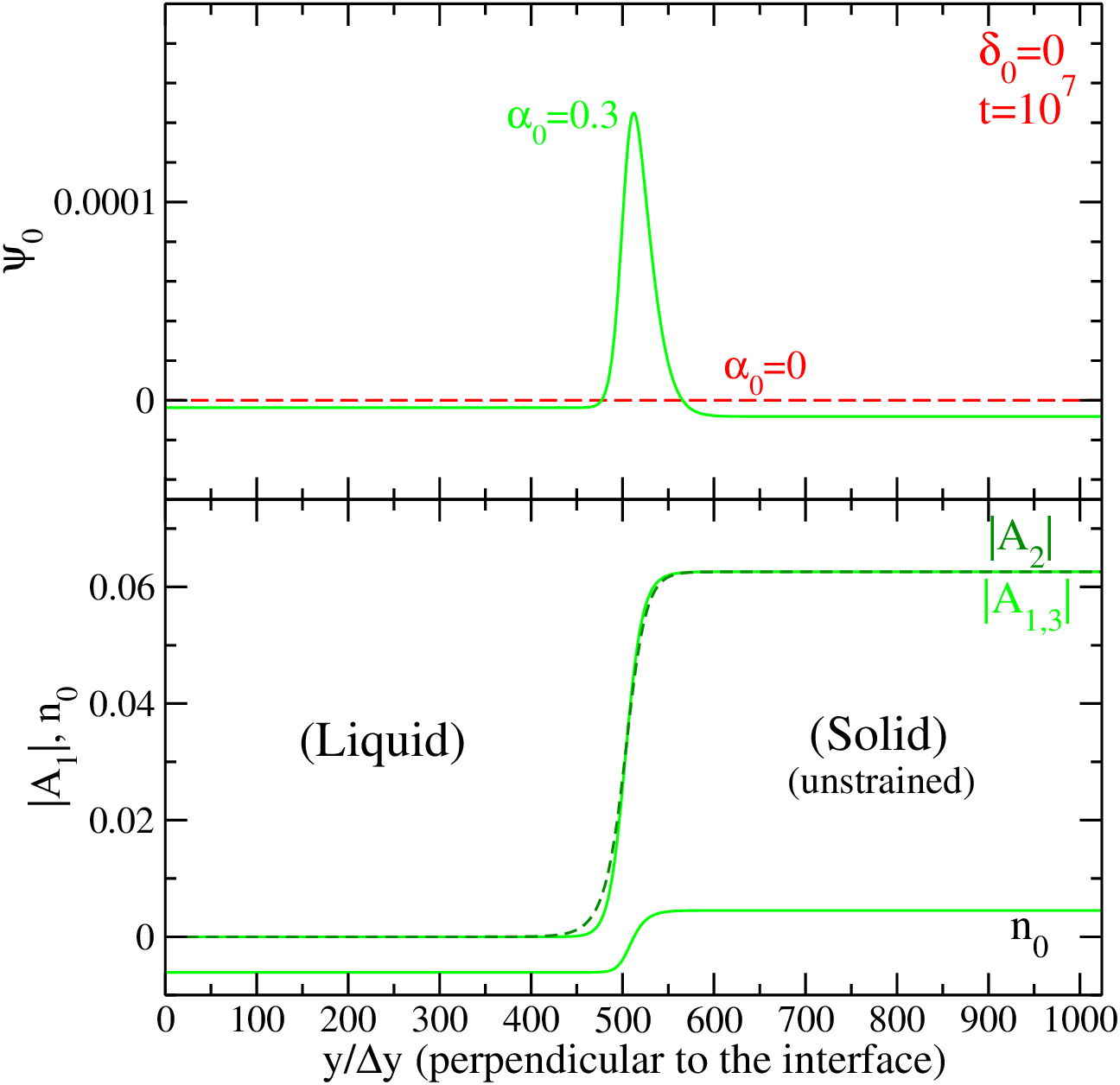}}
\caption{Liquid-solid coexistence profiles for unstrained solid layer
  (with $\delta_0=0)$ and $\alpha_0=0$ and $0.3$. The other parameters
  are the same as Fig. \ref{fig:slPeq}. In the lower panel the $|A_j|$
  and $n_0$ profiles overlap for $\alpha_0=0$ and $0.3$.}
\label{fig:sl0eq}
\end{figure}

It is interesting to note that the non-homogeneous compositional profile
can also be found in liquid-solid heterostructures with nonzero
$\alpha_0$ and no misfit strain (i.e., $\delta_0=0$, $\psi_0=0$, and
$\alpha_0=0.3$, as in Fig. \ref{fig:sl0eq}). A slight enrichment of
larger atoms A is observed on the surface of unstrained solid, showing
as a ``peak'' (with $\psi_0 \sim 1.5 \times 10^{-4}$) at the
compositional interface in Fig. \ref{fig:sl0eq}. Note that this
phenomenon of weak surface segregation persists in the equilibrium or
stationary configuration (as tested up to $t=10^7$), and is caused by
unequal atomic sizes of alloy components. Due to the conservation of 
the $\psi_0$ field and the appearance of concentration ``peak'' at
interface, the bulk values of concentration field $\psi_0$ in both
liquid and solid regions deviate from the 0 value in the corresponding
phase diagram, as mediated by the alloy diffusion process. We find
that this deviation is a result of finite size effect: The deviation
decreases with increasing system size, as confirmed in our simulations
of $L_y=1024 \Delta y$, $2048 \Delta y$ and $8192 \Delta y$. Thus in
the thermodynamic limit (with $L_y \rightarrow \infty$) $\psi_0=0$ is
expected in the liquid and solid bulks, consistent with the
equilibrium phase diagram for unstrained systems. On the other hand,
the effect of surface enrichment would be preserved, as we have
observed in simulations of various system sizes.

\subsubsection{Coherent strained layer growth and front motion}

\begin{figure}
\centerline{
\includegraphics[height=3.3in]{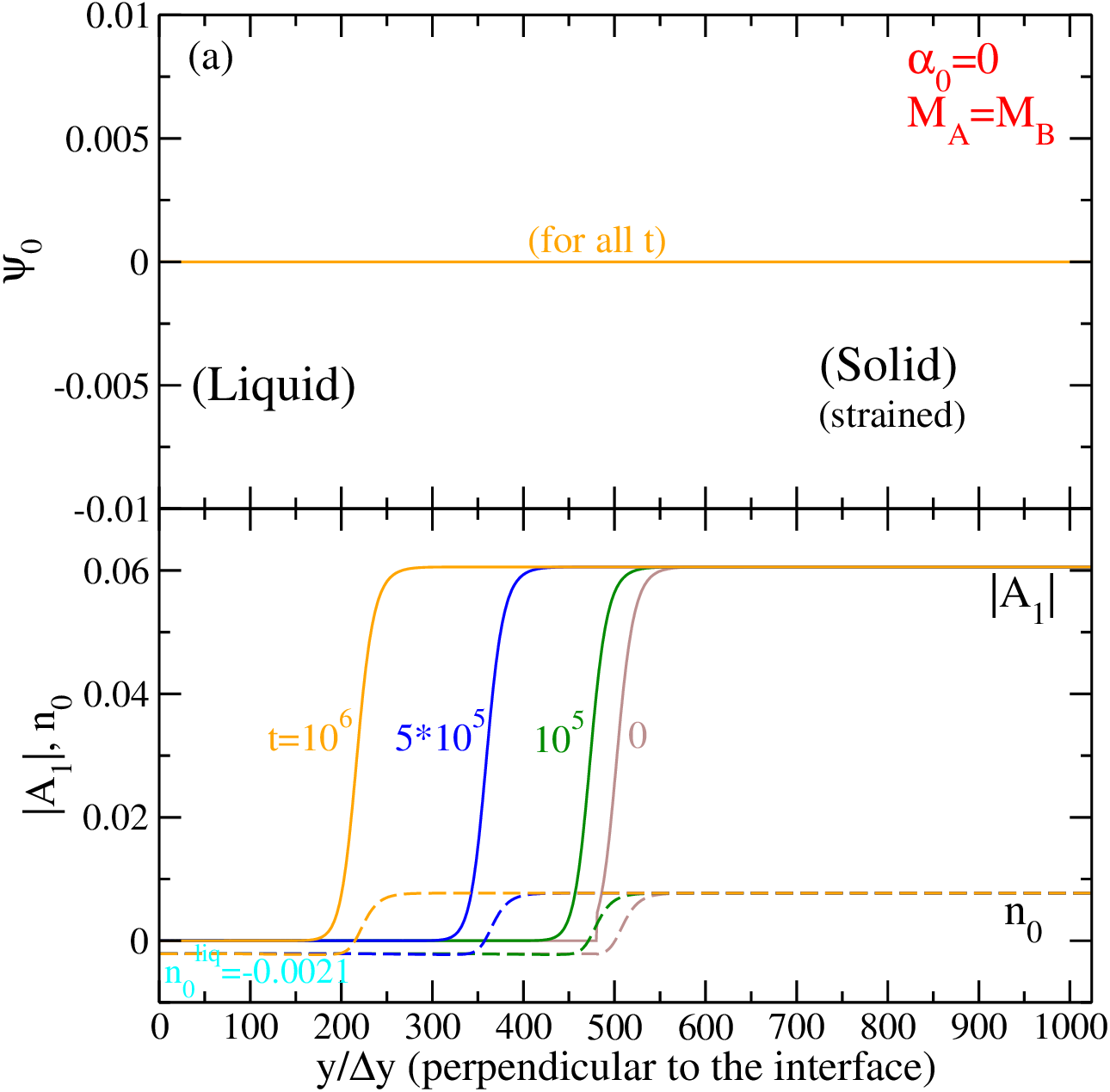} \hskip 5pt
\includegraphics[height=3.3in]{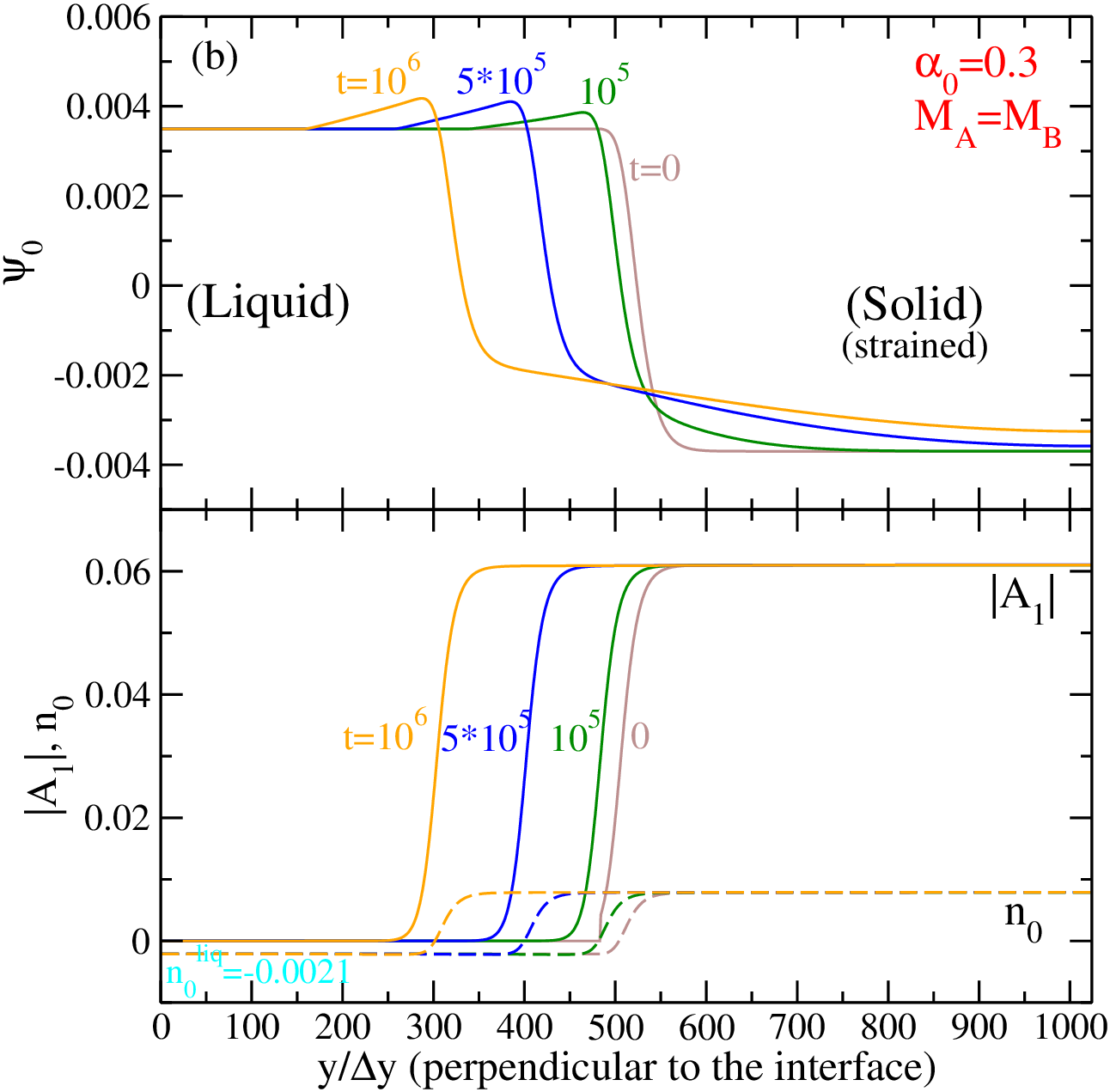}}
\caption{Growth of strained solid layer from a liquid-solid initial
  configuration, with $\alpha_0=0$ (a) and $0.3$ (b) and equal
  mobility $M_A=M_B$. The parameters are the same as the
  corresponding liquid-solid coexistence state given in
  Fig. \ref{fig:slPeq}, except for $n_0=-0.0021$ in the liquid
  region where a constant flux boundary condition is set up.}
\label{fig:sl_f}
\end{figure}

To simulate the process of strained layer growth encountered in
most experiments, we start from a
liquid-solid(strained) coexisting configuration and let the liquid
solidify, leading to a growing front of the strained
solid layer (as shown in Fig. \ref{fig:sl_f}). The initial condition
is set as the liquid-solid coexistence profiles given in
Fig. \ref{fig:slPeq}, with only $n_0$ in liquid changed to 
$n_0^{\rm liq}=-0.0021$ to initialize the solidification and
growth while all others (including concentration $\psi_0$ and
amplitudes $A_j$) being kept the same as the coexistence
condition. The growth rate of the strained layer can be controlled by
the setting of liquid $n_0^{\rm liq}$, i.e., its deviation from the
equilibrium or coexistence value. A boundary condition of constant
flux is kept in the liquid region (with distance $100 \Delta y$ beyond
the moving interface).

The growth process is shown in Fig. \ref{fig:sl_f}, for equal
mobility $M_A=M_B$, $5\%$ misfit strain for solid layer, and up to
$t=10^6$. The liquid-solid front moves smoothly for both $\alpha_0=0$
and $0.3$, as seen from the amplitude and $n_0$ profiles in the
figure. For $\alpha_0=0$, the concentration $\psi_0$ in both liquid
and solid layers remains uniform at the initial value 0, as in the equilibrium
state. However, the results for $\alpha_0=0.3$ show a phenomenon of
composition overshooting at the growth front of strained solid (see
Fig. \ref{fig:sl_f}b). Such overshooting effect reveals as the
increase of $\psi_0$ (i.e., more A or less B atoms) around the
interface, resulting in the phenomenon of surface enrichment: The A
atoms (with larger atomic size for $\alpha_0>0$) are segregated on the
solid surface with compressive strain.
As time increases, such variation of alloy concentration
will propagate into the bulk of solid layer as a result of atomic
diffusion (note that the concentration of liquid bulk remains
unchanged due to the constant flux boundary condition).

\begin{figure}
\centerline{
\includegraphics[height=3.3in]{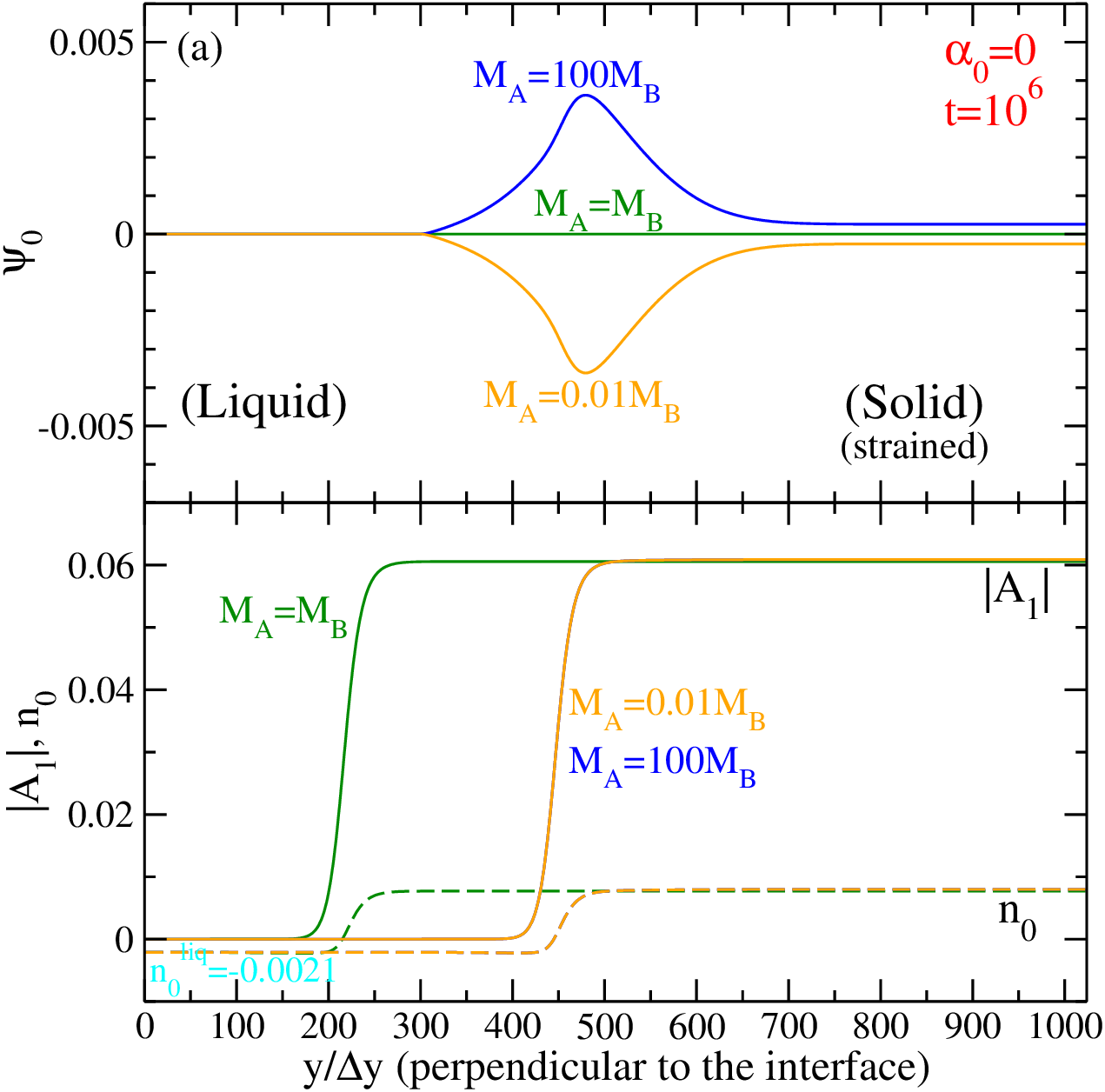} \hskip 5pt
\includegraphics[height=3.3in]{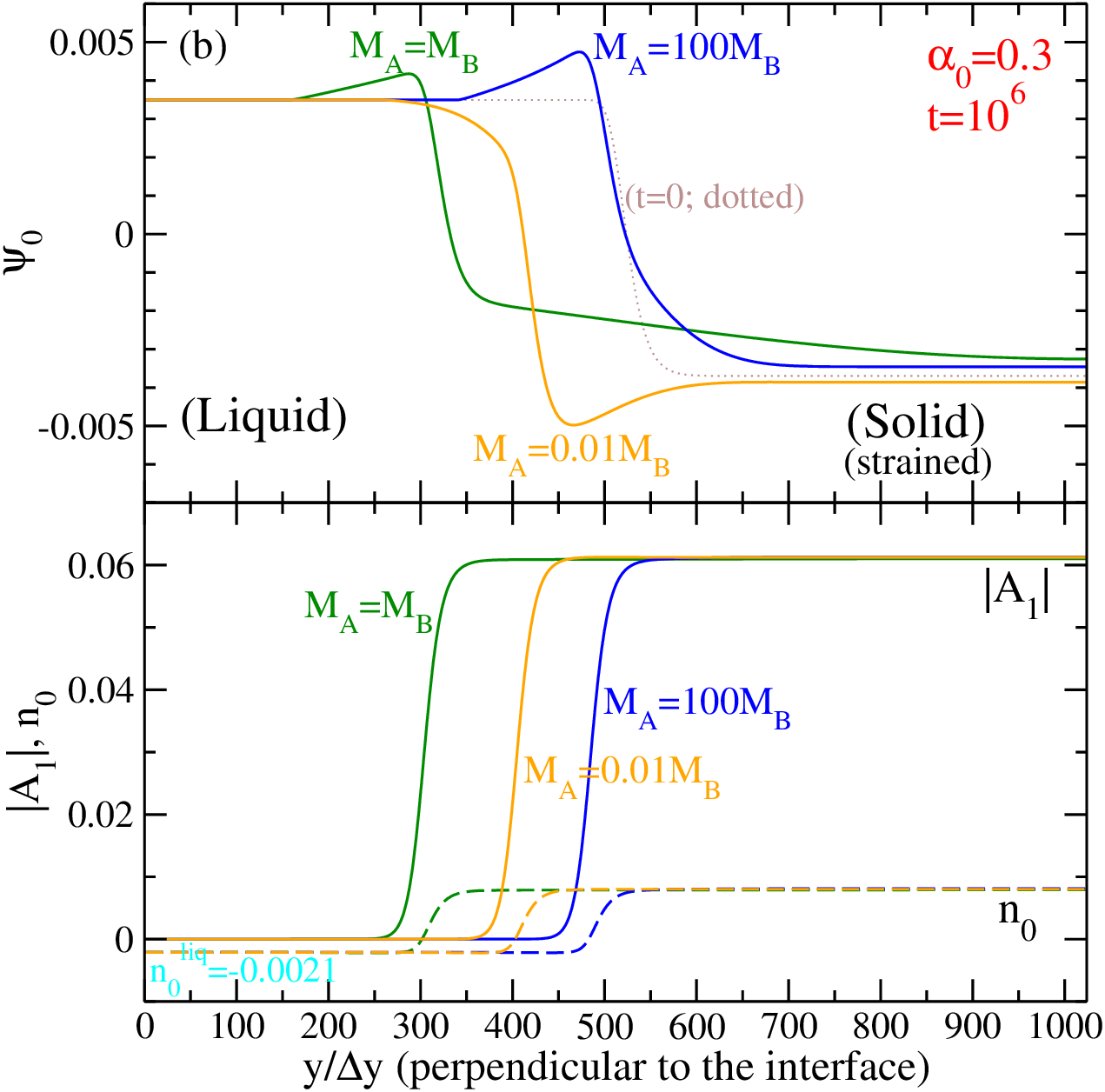}}
\caption{Growth of strained solid layer from a liquid-solid initial
  configuration, with $\alpha_0=0$ (a) and $0.3$ (b), 
  mobilities $M_A=M_B$, $M_A=100 M_B$, and $M_A=M_B/100$, and time
  $t=10^6$. Other parameters are the same as those in
  Fig. \ref{fig:sl_f}. In (a) the $|A_j|$ and $n_0$ profiles for
  $M_A=100 M_B$ and $M_A=M_B/100$ overlap.} 
\label{fig:sl_f_m}
\end{figure}

Figure \ref{fig:sl_f_m} shows that the mobility disparity between
different alloy components plays an important role on this
overshooting effect. Atoms with larger mobility will accumulate on
the surface, even with $\alpha_0=0$. As seen in the concentration
profile of Fig. \ref{fig:sl_f_m}a, a peak of larger (or smaller)
$\psi_0$ appears around the liquid-solid interface for $M_A > M_B$ (or
$M_A < M_B$), while no overshooting is observed in the case of equal
mobility. For nonzero $\alpha_0$ (Fig. \ref{fig:sl_f_m}b), the
effect of surface enrichment of A atoms will be enhanced when $M_A>
M_B$, while when $M_A < M_B$ the B atom enrichment
is observed at large enough time. 

Another effect of mobility difference presented in
Fig. \ref{fig:sl_f_m} is the change of solid layer growth rate or
front moving speed. For large disparity of atomic mobility between A
and B components, one of the components moves much slower compared to
the other one and thus would hinder the atomic diffusion process. This
leads to a slower motion of interface, as seen in
Fig. \ref{fig:sl_f_m}. Thus we can expect that
in the limit of $M_A/M_B \gg 1$ (or $M_A/M_B \ll 1$), B (or A)
atoms would be almost immobile compared to A (or B) and hence would
pin the interface location, resulting in a frozen front. This has been
incorporated in the amplitude equations developed above: When $m=\pm 1$
(with $m=(M_A-M_B)/(M_A+M_B)$ as defined in Eq. (\ref{eq:m_g0_v})),
Eq. (\ref{eq:Aj0}) yields $dA_j/dt=0$, a frozen amplitude profile.
Furthermore, the concentration profile is symmetric with respect to
the sign of $m$ (i.e., $M_A/M_B>1$ vs. $<1$) for $\alpha_0=0$, as
shown in Fig. \ref{fig:sl_f_m}a for $M_A/M_B=100$ and $10^{-2}$ which
yield the same front moving rate and the same $A_j$ and $n_0$ profiles. 
The situation for nonzero $\alpha_0$ (different atomic sizes) is more
complicated. In our calculations of Fig. \ref{fig:sl_f_m}b with
$\alpha_0=0.3$ and $5\%$ compressive misfit, the liquid-solid
coexisting profile yields $\psi_0>0$ (A-rich) in the liquid region and
$<0$ (B-rich) in the solid layer (see also Fig. \ref{fig:slPeq}b). 
When $M_A=100 M_B$, the segregation of fast A atoms around the
interface would tend to hinder the growth of B-rich solid layer, while
for $M_A=M_B/100$ the accumulation of fast B atoms will naturally be
accompanied by the expansion of solid region, resulting in a
faster solid growth.

The composition overshooting effect presented here and the associated
surface enrichment phenomenon can be viewed as a result of 
interface intermixing process via atomic interdiffusion and mass
transport of alloy components, showing as the vertical phase
separation or segregation in the liquid-solid interface region. Such
process of vertical separation has also been found in 2D
simulations of binary PFC equations \cite{re:elder07}, where the
component of greater size or larger mobility was found to accumulate
near undulated solid surface in a liquid/substrate epitaxial
system. Importantly, the results shown here are consistent with recent
experimental observations of surface or interface segregation
phenomenon in alloy heterostructures, particularly in semiconductor
epitaxial layers. Most experiments focus
on III-V or group IV heteroepitaxial films, with typical systems
including InGaAs/GaAs(001) (with In enrichment or segregation
\cite{re:moison89,re:walther01,re:cederberg07}), Ge(SiGe)/Si(001)
(with Ge segregation \cite{re:walther97,re:denker05}),
and multilayers or superlattices of InP/InGaAs (with excess InAs at
the interface \cite{re:gerling01}), GaAs/GaSb (with Sb segregation and
Sb-As exchange and intermixing \cite{re:dorin03}), GaAs/InAs
(with In segregation \cite{re:pearson04}), etc. In these experimental
systems the segregation or enrichment effect involves the coupling of
various factors of different atomic size (nonzero $\alpha_0$), misfit
strain, and unequal mobility of alloy components (e.g., $M_{\rm Ge} >
M_{\rm Si}$ and $M_{\rm In} > M_{\rm Ga}$), each of which has been
identified in our analysis given above.

\section{Conclusions}

In this paper we have furthered the development of the
phase-field-crystal methodology by systematically deriving the PFC
dynamic model equations from dynamical density functional theory
(DDFT) and completing the derivation of the corresponding amplitude equation formalism. 
A truncation of the DFT free energy functional up to three-point direct
correlation functions has been used, and the dynamics derived from
DDFT has been further simplified through lowest order approximations
via a simple scale analysis to obtain the PFC equations, for both
single-component and binary alloy systems. 
For the binary PFC model, the corresponding amplitude
equations (both deterministic and stochastic) have been established
via a hybrid multiple-scale approach, which describe large or ``slow''
scale dynamics of structural and compositional profiles based on the
underlying crystalline state. Compared to other recent developments
which have mainly focused on the evolution of complex structural amplitudes
and concentration field, this work presents results 
that incorporate the new effects of mobility
difference between alloy components, the coupling to zero-mode average
atomic density, and also noise dynamics.
Although the results of amplitude equations that we derive are for 2D
hexagonal crystalline state, they can be extended to 3D bcc or fcc
structures by following a procedure similar to the one developed here
and adopting the corresponding basic wavevectors (see also
Ref. \cite{re:elder10}).

This amplitude equation formalism for binary PFC has been applied to
identifying the mechanisms and parameter coupling during the process
of surface segregation and alloy intermixing. Both liquid-solid and
liquid-solid-solid epitaxial heterostructures have been examined,
including morphological and compositional profiles.
We find that the effect of concentration segregation on solid surface is
controlled by material parameters such as the disparity of atomic size
and mobility between different alloy components and misfit strain
in solid layers. In the cases of nonzero solute expansion coefficient
or unequal atomic mobility, an effect of composition overshooting around
liquid-solid interface is obtained during strained layer growth, 
corresponding to vertical phase separation or segregation in the interface
region. These results are consistent with recent experimental findings
in heteroepitaxial systems, particularly the phenomenon of surface or
interface segregation showing as the enrichment of one of the alloy
species as compared to the bulk phase. This sample application of the
amplitude equation formalism developed here has further illustrated
the features and advantages of the PFC methodology, particularly in
terms of modeling and understanding complex material phenomena
involving spacial and temporal scales of experimental relevance.

\begin{acknowledgments}
Z.-F.H. acknowledges support from the National Science Foundation
(NSF) under Grant No. CAREER DMR-0845264. K.R.E. acknowledges support
from NSF under Grant No. DMR-0906676. N.P. acknowledges support from
the National Science and Engineering Research Council of Canada.
\end{acknowledgments}

\appendix
\section{Alternative derivations of binary PFC dynamics via DDFT}
\label{append:altern}

\subsection{Alternative derivation I}

In the following we provide an alternative derivation procedure for
PFC dynamics, including two steps:
i) directly use the original free energy functional (\ref{eq:F_AB})
and the DDFT equations (\ref{eq:ddft_AB}) to obtain the expressions of
$\partial \rho_{A(B)} / \partial t$, and then ii) derive the dynamics of
$n$ and $\psi$ through Eq. (\ref{eq:ndN_rho}), instead of using
Eqs. (\ref{eq:F_ndN}) and (\ref{eq:D_12}) as in
Sec. \ref{sec:ddft_binary}.

Define $n_A = (\rho_A - \rho_l^A)/\rho_l$ and $n_B = (\rho_B -
\rho_l^B)/\rho_l$, and hence the free energy functional
(\ref{eq:F_AB}) can be rewritten as (using the expansion
(\ref{eq:C_expan}))
\begin{eqnarray}
&\Delta {\cal F}/\rho_l k_BT =& \int d{\bm r} \left \{
\Delta \rho_l^A \left ( 1+\frac{n_A}{\Delta \rho_l^A} \right ) 
\ln \left ( 1+\frac{n_A}{\Delta \rho_l^A} \right )
+ \Delta \rho_l^B \left ( 1+\frac{n_B}{\Delta \rho_l^B} \right ) 
\ln \left ( 1+\frac{n_B}{\Delta \rho_l^B} \right )
\right. \nonumber\\
&& -(n_A+n_B) + \frac{\rho_l}{2} \left [ n_A \left ( 
   \hat{C}_2^{AA} \nabla^2 + \hat{C}_4^{AA} \nabla^4 \right ) n_A
 + n_B \left (\hat{C}_2^{BB} \nabla^2 +
  \hat{C}_4^{BB} \nabla^4 \right ) n_B \right. \nonumber\\
&& \left. + 2n_A \left ( \hat{C}_2^{AB} \nabla^2 +
  \hat{C}_4^{AB} \nabla^4 \right ) n_B 
+ \hat{C}_0^{AA} n_A^2 + \hat{C}_0^{BB} n_B^2 
+ 2\hat{C}_0^{AB} n_An_B \right ] \nonumber\\
&& \left. + \frac{\rho_l^2}{6} \left [ \hat{C}_0^{AAA} n_A^3
+ \hat{C}_0^{BBB} n_B^3 + 3\hat{C}_0^{AAB} n_A^2n_B
+ 3\hat{C}_0^{ABB} n_An_B^2 \right ] \right \},
\label{eq:F_nAB}
\end{eqnarray}
where $\Delta \rho_l^A = \rho_l^A/\rho_l$ and $\Delta \rho_l^B =
\rho_l^B/\rho_l$. From the DDFT equations (\ref{eq:ddft_AB}) we can
obtain the PFC equations for A \& B components respectively, i.e.,
\begin{eqnarray}
&\partial n_A /\partial t =& M_A k_B T \left \{ \nabla^2 \left
    [ \left ( 1+\rho_l^A \hat{C}_0^{AA} \right ) n_A 
    + \rho_l^A \left ( \hat{C}_2^{AA} \nabla^2 +
      \hat{C}_4^{AA} \nabla^4 \right ) n_A 
+ \frac{1}{3} \rho_l^2 \hat{C}_0^{AAA} n_A^3
     \right. \right. \nonumber\\
&& \left. + \frac{\rho_l}{2} \left ( \hat{C}_0^{AA} 
      + \rho_l^A \hat{C}_0^{AAA} \right ) n_A^2 \right ]
+ \rho_l {\bm \nabla} \cdot \left [ n_A \left ( \hat{C}_2^{AA}
    \nabla^2 + \hat{C}_4^{AA} \nabla^4 \right ) {\bm \nabla} n_A
\right ] \nonumber\\
&& + \rho_l {\bm \nabla} \cdot \left [ (n_A + \Delta \rho_l^A) 
{\bm \nabla} \left [ \left ( \hat{C}_0^{AB} + \hat{C}_2^{AB} \nabla^2 +
    \hat{C}_4^{AB} \nabla^4 \right ) n_B \right. \right. \nonumber\\
&& \left. \left. \left. 
+ \frac{\rho_l}{2} \left ( 2\hat{C}_0^{AAB} n_An_B +
  \hat{C}_0^{ABB} n_B^2 \right ) \right ] \right ] \right \},
\label{eq:pfc_nA}
\end{eqnarray}
\begin{eqnarray}
&\partial n_B /\partial t =& M_B k_B T \left \{ \nabla^2 \left
    [ \left ( 1+\rho_l^B \hat{C}_0^{BB} \right ) n_B 
    + \rho_l^B \left ( \hat{C}_2^{BB} \nabla^2 +
      \hat{C}_4^{BB} \nabla^4 \right ) n_B
+ \frac{1}{3} \rho_l^2 \hat{C}_0^{BBB} n_B^3
     \right. \right. \nonumber\\
&& \left. + \frac{\rho_l}{2} \left ( \hat{C}_0^{BB} 
      + \rho_l^B \hat{C}_0^{BBB} \right ) n_B^2 \right ]
+ \rho_l {\bm \nabla} \cdot \left [ n_B \left ( \hat{C}_2^{BB}
    \nabla^2 + \hat{C}_4^{BB} \nabla^4 \right ) {\bm \nabla} n_B
\right ] \nonumber\\
&& + \rho_l {\bm \nabla} \cdot \left [ (n_B + \Delta \rho_l^B) 
{\bm \nabla} \left [ \left ( \hat{C}_0^{AB} + \hat{C}_2^{AB} \nabla^2 +
    \hat{C}_4^{AB} \nabla^4 \right ) n_A \right. \right. \nonumber\\
&& \left. \left. \left. 
+ \frac{\rho_l}{2} \left ( 2\hat{C}_0^{ABB} n_An_B +
  \hat{C}_0^{AAB} n_A^2 \right ) \right ] \right ] \right \}.
\label{eq:pfc_nB}
\end{eqnarray}
Note that $n=n_A+n_B$ and $\psi = [ (\rho_l^A-\rho_l^B) + \rho_l
(n_A-n_B)] / [\rho_l (1+n)]$, and hence equivalent to
Eq. (\ref{eq:ndN_rho}) we have
\begin{equation}
\frac{\partial n}{\partial t} = \frac{\partial n_A}{\partial t} 
+ \frac{\partial n_B}{\partial t}, \qquad
\frac{\partial \psi}{\partial t} = \frac{1}{1+n} 
\left [ (1-\psi) \frac{\partial n_A}{\partial t}
- (1+\psi) \frac{\partial n_B}{\partial t} \right ].
\label{eq:ndN_nAB}
\end{equation}
Substituting Eqs. (\ref{eq:pfc_nA}) and (\ref{eq:pfc_nB}) into
Eq. (\ref{eq:ndN_nAB}), and noting $n_A=(1+n)(1+\psi)/2-\Delta
\rho_l^A$, $n_B=(1+n)(1-\psi)/2-\Delta \rho_l^B$, 
$M_A = \rho_l (M_1 + M_2)$, and $M_B = \rho_l (M_1 - M_2$), we can derive
the binary PFC equations for $n$ and $\psi$, which are exactly the
same as Eqs. (\ref{eq:pfc_ndN}), (\ref{eq:D1}), and (\ref{eq:D2}).

\subsection{Alternative derivation II}

Another alternative derivation for PFC dynamics is to start with
Eqs. (\ref{eq:pfc_ndN})--(\ref{eq:D_12}) as already derived in
Sec. \ref{sec:ddft_binary}. Different from Sec. \ref{sec:ddft_binary},
in the formula (\ref{eq:D_12}) for ${\cal D}_1$ and ${\cal D}_2$ if
considering that the chemical potentials $\mu_n = \delta {\cal F} /
\delta n$ and $\mu_N = \delta {\cal F} / \delta \psi$ are slowly
varying quantities and retaining terms up to the lowest order, we have
\begin{equation}
{\cal D}_1 \simeq \nabla^2 \frac{\delta {\cal F}}{\delta n}, \qquad
{\cal D}_2 \simeq \nabla^2 \frac{\delta {\cal F}}{\delta \psi},
\label{eq:D12II}
\end{equation}
as used in the original PFC model \cite{re:elder07,re:elder10}. As
in the previous work, the logarithm terms in the free energy functional
(\ref{eq:F_ndN}) are expanded in a power series, yielding (up to 4th
order of $n$ and $\psi$)
\begin{eqnarray}
&\Delta {\cal F}/\rho_l k_BT =& \int d{\bm r} \left \{
\left [ \frac{1}{2} \psi^2 + \beta (\psi) \right ] n + \frac{1}{2}
B^{\ell}(\psi) n^2 + \frac{1}{3} \left ( \tau + \tilde{B}_1 \psi \right ) n^3
+ \frac{1}{4} v n^4 + \frac{1}{2} w \psi^2 
+ \frac{1}{3} \beta_3 \psi^3 + \frac{1}{4} u \psi^4 \right. \nonumber \\
&& \left. + \frac{1}{2} (1+n) \left ( 2B^x R^2 \nabla^2 + B^x R^4
\nabla^4 \right ) n + \frac{1}{2} K \left | {\bm \nabla}
[(1+n)\psi] \right |^2 + \frac{\kappa}{2} \left ( \nabla^2 [(1+n)\psi]
\right )^2 \right \},
\label{eq:F_ndN_II}
\end{eqnarray}
where $\tau = \tilde{B}_0 - 1/2$, $w =  1+\beta_2$, $v = u = 1/3$,
and other parameters (such as $B^{\ell}$, $\beta$, $K$, and
$\kappa$) are defined in Eq. (\ref{eq:parameters}).

Substituting Eq. (\ref{eq:F_ndN_II}) into (\ref{eq:D12II}), 
we find (up to 3rd order)
\begin{eqnarray}
&{\cal D}_1 =& \nabla^2 \left \{ -(B_0^x - B_0^{\ell}) n + B_0^x \left (R_0^2
  \nabla^2 + 1 \right )^2 n + (B_1^{\ell} \psi + B_2^{\ell} \psi^2) n +
(\tau + \tilde{B}_1 \psi) n^2 + v n^3 
\right. \nonumber\\
&& + \beta_0 \psi + \left ( \frac{1}{2} + \beta_1 \right ) \psi^2 
+ \beta_3 \psi^3  + \psi \left ( - K \nabla^2 + \kappa \nabla^4 
\right ) \left [ (1+n) \psi \right ] \nonumber\\
&& \left. + B_0^x \left ( \alpha_2 R_0^2
    \nabla^2 + \frac{\alpha_4}{2} R_0^4 \nabla^4 \right ) \left [
    (1+n) \psi \right ] + B_0^x \psi \left ( \alpha_2 R_0^2 \nabla^2 +
    \frac{\alpha_4}{2} R_0^4 \nabla^4 \right ) n \right \},\\
&{\cal D}_2 =& \nabla^2 \left \{ \beta_0 n + B_0^x (1+n) \left (
  \alpha_2 R_0^2 \nabla^2 + \frac{\alpha_4}{2} R_0^4 \nabla^4 
  \right ) n + \left [ (1+2\beta_1) \psi + 3\beta_3 \psi^2 \right ] n 
  + \frac{1}{2} \left ( B_1^{\ell} + 2B_2^{\ell} \psi \right ) n^2
  + \frac{1}{3} \tilde{B}_1 n^3 \right. \nonumber\\
&& \left. + w \psi + \beta_3 \psi^2 + u \psi^3 + (1+n) \left ( - K
    \nabla^2 + \kappa \nabla^4 \right ) \left [ (1+n) \psi \right ] 
\right \}.
\end{eqnarray}
As in Sec. \ref{sec:ddft_binary}, we can rescale the PFC equations
(using the same length and time scales as well as the scales for $n$
and $\psi$ fields). Following the scale analysis discussed at the
end of Sec. \ref{sec:ddft_binary}, to lowest order approximation we
can obtain the same simplified binary PFC equations given in
Eqs. (\ref{eq:pfc_npsi}) and (\ref{eq:D12}), albeit with different
forms of rescaled parameters $g_2 = g_0 \tau /B_0^x$ and
$v_1=g_0(1/2+\beta_1)/B_0^x$ (other parameters $g$, $w_0$, $\alpha_0$,
and $g_0$ are the same as those defined in Sec. \ref{sec:ddft_binary}
but with different value of $v$).

%\bibliographystyle{apsrev}
%\bibliography{../references}

\begin{thebibliography}{91}
\bibitem[{re:()}]{re:notes_pfc}
\bibinfo{note}{For the ideal-gas part $(1+n)\ln (1+n)$ in the free energy
  (\ref{eq:F_C3}), it contributes to a term $\nabla^2 n$ in the dynamics ${\bm
  \nabla} \cdot \left [ (1+n) {\bm \nabla} \delta {\cal F} / \delta n \right ]$
  of Eq. (\ref{eq:ddft_n}). It is straightforward to show that this can also be
  obtained via first expanding $(1+n)\ln(1+n) = n + n^2/2 - n^3/6 + n^4/12 -
  \cdots$ as a power series of $n$, similar to the previous work of original
  PFC model \cite{re:elder07,re:jaatinen09} but without truncating at any given
  order, and then substituting into the functional derivative.}
\end{thebibliography}
%\input{bpfc_ampl.bbl}

\end{document}